\documentclass[aps,prd,onecolumn,showpacs,superscriptaddress,nofootinbib]{revtex4-1}
\expandafter\let\csname equation*\endcsname\relax
\expandafter\let\csname endequation*\endcsname\relax
\usepackage{amsmath}
\usepackage{amssymb}
\usepackage{amsbsy}
\usepackage{color}
\usepackage{graphicx}
\usepackage{subfigure}
\usepackage{float}
\usepackage{dcolumn}
\usepackage{footnote}
\usepackage{url}
\usepackage{rotating}
\usepackage[normalem]{ulem}
\usepackage{natbib}

\newcommand\apjl{The Astrophysical Journal Letters}
\newcommand\AIAA{American Institute of Aeronautics and Astronautics}
\newcommand\caa{Chinese Astronomy and Astrophysics}

\DeclareMathOperator*{\argmin}{arg\,min}

\begin{document}

\title{Orbital effects on time delay interferometry for TianQin}

\author{Ming-Yue Zhou}
\affiliation{MOE Key Laboratory of Fundamental Physical Quantities Measurements,
Hubei Key Laboratory of Gravitation and Quantum Physics, PGMF, 
Department of Astronomy, and 
School of Physics, Huazhong University of Science and Technology,
Wuhan 430074, China}

\author{Xin-Chun Hu}
\affiliation{MOE Key Laboratory of Fundamental Physical Quantities Measurements,
Hubei Key Laboratory of Gravitation and Quantum Physics, PGMF, 
Department of Astronomy, and 
School of Physics, Huazhong University of Science and Technology,
Wuhan 430074, China}

\author{Bobing Ye}
\affiliation{TianQin Research Center for Gravitational Physics \& School of Physics and Astronomy, 
Sun Yat-sen University (Zhuhai Campus), Zhuhai 519082, China}

\author{Shoucun Hu}
\affiliation{CAS Key Laboratory of Planetary Sciences, Purple Mountain Observatory, Chinese Academy of Sciences, Nanjing 210033, China}
\affiliation{CAS Center for Excellence in Comparative Planetology, Hefei 230026, China}

\author{Dong-Dong Zhu}
\affiliation{MOE Key Laboratory of Fundamental Physical Quantities Measurements,
Hubei Key Laboratory of Gravitation and Quantum Physics, PGMF, 
Department of Astronomy, and 
School of Physics, Huazhong University of Science and Technology,
Wuhan 430074, China}

\author{Xuefeng Zhang}
\affiliation{TianQin Research Center for Gravitational Physics \& School of Physics and Astronomy, 
Sun Yat-sen University (Zhuhai Campus), Zhuhai 519082, China}

\author{Wei Su}
\affiliation{MOE Key Laboratory of Fundamental Physical Quantities Measurements,
Hubei Key Laboratory of Gravitation and Quantum Physics, PGMF, 
Department of Astronomy, and 
School of Physics, Huazhong University of Science and Technology,
Wuhan 430074, China}

\author{Yan Wang}
\email{ywang12@hust.edu.cn}
\affiliation{MOE Key Laboratory of Fundamental Physical Quantities Measurements,
Hubei Key Laboratory of Gravitation and Quantum Physics, PGMF, 
Department of Astronomy, and 
School of Physics, Huazhong University of Science and Technology,
Wuhan 430074, China}

\date{\today}

\begin{abstract}

The proposed space-borne laser interferometric gravitational wave (GW) observatory TianQin adopts  
a geocentric orbit for its nearly equilateral triangular constellation formed by three identical drag-free satellites. 
The geocentric distance of each satellite is $\approx 1.0 \times 10^{5} ~\mathrm{km}$, 
which makes the armlengths of the interferometer be $\approx 1.73 \times 10^{5} ~\mathrm{km}$. 
It is aimed to detect the GWs in $0.1 ~\mathrm{mHz}-1 ~\mathrm{Hz}$.
For space-borne detectors, the armlengths are unequal and change continuously
which results in that the laser frequency noise is nearly $7-8$ orders of magnitude
higher than the secondary noises (such as acceleration noise, optical path noise, etc.).
The time delay interferometry (TDI) that synthesizes virtual interferometers
from time-delayed one-way frequency measurements has been proposed to suppress
the laser frequency noise to the level that is comparable or below the secondary noises.
In this work, we evaluate the performance of various data combinations for both first- 
and second-generation TDI based on the five-year numerically optimized orbits of
the TianQin's satellites which exhibit the actual rotating and flexing of the constellation. 
We find that the time differences of symmetric interference paths of the data
combinations are $\sim 10^{-8}$~s for the first-generation TDI and  $\sim 10^{-12}$~s 
for the second-generation TDI, respectively. While the second-generation TDI is
guaranteed to be valid for TianQin, the first-generation TDI is possible to be
competent for GW signal detection with improved stabilization of the laser frequency noise in the concerned GW frequencies.

\end{abstract}

\maketitle

\section{Introduction}\label{sec:intro} 

The detection of gravitational waves (GWs) from the coalescence of a stellar-mass black 
hole binary (GW150914) by advanced LIGO detectors \cite{2016PhRvL.116f1102A} has opened  
up the era of observational GW astronomy. 
During the first two observing runs (O1 and O2) \cite{2019PhRvX...9c1040A} 
and the first half of O3 \cite{2020arXiv201014527A} 
of advanced LIGO and advanced Virgo, more than 50 compact binary coalescences, 
including the first observation of binary neutron star inspiral (GW170817) \cite{2017PhRvL.119p1101A}, 
have been detected. 
Both advanced LIGO and advanced Virgo will reach their design sensitivities in the coming 
years which can boost the detection of GW events to higher rate. 
In addition, the underground cryogenic GW telescope KAGRA \cite{Somiya_2012} 
has recently joined in the advanced ground-based detector network.

The observational window of GW astronomy will be broadened to millihertz range 
(0.1 mHz--1 Hz) by the proposed space-borne laser interferometers, such as  
Laser Interferometer Space Antenna (LISA) \cite{2017arXiv170200786A}, 
TianQin \cite{2016CQGra..33c5010L}, DECIGO \cite{2011CQGra..28i4011K}, 
ASTROD-GW \cite{1998grco.conf..309N}, gLISA \cite{Tinto2015gLISA}, 
Taiji (ALIA descoped) \cite{2015JPhCS.610a2011G} and BBO \cite{Cutler2006BBO}. 
Among these, LISA has been comprehensively studied and developed 
for more than three decades \cite{LISA_2000}. 
In 2017, it has been selected as ESA's L-3 mission Cosmic Vision programme 
with the theme of ``the Gravitational Universe". 
LISA is scheduled for launch in early 2030s and will be operated concurrently with 
ESA's next generation Advanced Telescope for High ENergy Astrophysics (Athena). 
The latter will tremendously enhance the follow-up X-ray 
observation of the electromagnetic counterparts 
of LISA's GW source candidates \cite{Athena-LISA_2019}. 
The successful flight of LISA Pathfinder has demonstrated the feasibility of the key technologies, 
such as gravity reference system and space laser interferometry, to be implemented 
in LISA \cite{PhysRevLett.116.231101, 2018PhRvL.120f1101A}. 
The recent GRACE Follow-on mission has demonstrated the technologies to be used in 
the laser ranging interferometer (LRI) of LISA \cite{2019PhRvL.123c1101A}.

Similar to LISA, TianQin is comprised of three identical drag-free satellites that form a nearly 
equilateral triangular constellation  \cite{2016CQGra..33c5010L}. 
Each pair of satellites is linked by two one-way infrared laser 
beams which can be used, together with the intra-satellite laser links, 
to synthesize up to three Michelson interferometers. 
Distinct from LISA, TianQin adopts a geocentric orbit with an altitude of ${10^5}$ km 
from the geocenter, hence the armlength of each side of the 
triangle is approximately $1.73 \times {10^5}$ km. 
The detector plane formed by the three satellites faces to the galactic white dwarf binary 
RX J0806+1527 \cite{Strohmayer2008} (see Fig.~\ref{fig:1}). 
The guiding center of the constellation coincides with the geocenter, and 
the period of each satellite orbiting around the Earth is 3.65 days. 
Analytic approximation of the orbit coordinates for each satellite and 
the strain output of a Michelson interferometer for arbitrary incoming GWs 
are studied in \cite{Hu2018}. 
A series of study on TianQin's orbit and constellation, including constellation 
stability optimization \cite{2019IJMPD..2850121Y}, orbital orientation and 
radius selection \cite{2020IJMPD..2950056T}, eclipse avoidance \cite{2020arXiv201203269Y}, 
and the Earth-Moon's gravity disturbance evaluation \cite{2020arXiv201203264Z}, have been conducted. 
The recent progresses in the investigations of both science case and technological realization for TianQin 
can be found in \cite{2019PhRvD..99l3002F, 2019PhRvD.100d3003W, 2019PhRvD.100d4036S, 2019PhRvD.100h4024B, PhysRevD.101.103027, PhysRevD.102.063016, PhysRevD.102.063021,
2020CQGra..37r5013L,2020CQGra..37k5005Y,Su2020A,2020PlST...22k5301L}. 
A brief summary can be found in \cite{Mei2020ptep}.

The GW sources in the mHz frequency regime are rich, 
which include coalescing supermassive black hole binaries (SMBHBs), 
ultracompact binaries in the Galaxy, extreme mass ratio inspirals (EMRIs), 
stochastic GW background, etc \cite{2009LRR....12....2S, Hu2017}. 
For TianQin, preliminary studies on the detection rate of the SMBHBs 
\cite{2019PhRvD.100d3003W}, the associated parameter estimation 
accuracy based on the inspiral signals \cite{2019PhRvD..99l3002F}, and 
testing the no-hair theorem \cite{2019PhRvD.100d4036S} and 
constraining the modified gravity \cite{2019PhRvD.100h4024B}
with the post-merger ringdown signals have been carried out. 
The prospects for detecting galactic double white dwarfs \cite{PhysRevD.102.063021}, 
EMRIs \cite{PhysRevD.102.063016} and stellar-mass black hole binaries \cite{PhysRevD.101.103027} 
with TianQin have been investigated.

Unlike the ground-based interferometers, the armlengths of a space-borne interferometer are unequal and varying in time. 
Therefore, the common mode laser frequency (or phase) noise, which is $7-8$ orders of magnitude higher than 
the secondary noises (such as optical path noise and acceleration noise), cannot be canceled out 
at the phasemeter where the delay lines from different paths are interfered. 
The time delay interferometry (TDI) has been proposed to suppress the laser frequency noise for 
space-borne interferometers to the level that is comparable or below the secondary noises, 
while conserving the GW content in the data stream, by synthesizing 
virtual interferometers with time-delayed one-way frequency measurements 
\cite{1987ApJ...318..536A,1999ApJ...527..814A, 2000PhRvD..62d2002E,1999PhRvD..59j2003T, 
2001PhRvD..63b1101T, 2002PhRvD..65h2003T, 2003PhRvD..68f1303S, 2004PhRvD..69h2001T, 
2004PhRvD..70j2003N}. 
The first-generation TDI is devised to remove the laser frequency noise in a static detector configuration. 
Multiple data combinations have been found which include the six-pulse combinations $(\alpha,\beta,\gamma,\zeta)$, 
the eight-pulse combinations, such as unequal-arm Michelson $(X,Y,Z)$, Relay $(U,V,W)$, 
Monitor $(E,F,G)$ and Beacon $(P ,Q,R)$ (see examples in Fig.~\ref{fig:4}) 
\cite{1999ApJ...527..814A, 2005LRR.....8....4T}, 
and the optimal combinations $(A,E,T)$ \cite{2002PhRvD..66l2002P}.

All data combinations are aimed to make the lengths of the two symmetric interference 
paths in the synthesized interferometric measurements nearly equal. 
However, in the reality, this equality cannot be exactly satisfied due to the orbital dynamics of each satellite. 
The second-generation TDI has been proposed to further account for the rotation of the constellation 
and the linear variation of armlengths \cite{2003PhRvD..68f1303S, 2004PhRvD..69h2001T,2003CQGra..20.4851C}, 
which improves the length equality by judiciously 
splicing the first-generation interference paths \cite{2005PhRvD..72d2003V}. 
This results in more data combinations than the first-generation TDI.

The result from \textit{Synthetic LISA}, based on the analytic approximation of LISA spacecraft's 
orbits \cite{Dhurandhar_2005}, shows that the time differences of the 
symmetric interference paths are $10^{-6}$ s and $10^{-10}$~s 
for the first- and second-generation TDI, hence the latter must be adopted 
in LISA data analysis to comfortably cancel out the laser phase noises \cite{2005PhRvD..71b2001V}. 
This is further confirmed by the detailed numerical simulations, based on the orbits optimized
with CGC 2.7 ephemeris \cite{CGC2_2002}, 
of various first- and second-generation TDI data combinations 
for (e)LISA \cite{2013CQGra..30f5011W, 2013AdSpR..51..198D}. 
Similar investigations have also been conducted for ASTROD-GW \cite{2012ChA&A..36..211W, 2015ChPhB..24e9501W} 
and Taiji \cite{2020arXiv200805812W}.

In this work, we simulate the time differences of the symmetric interference paths of 
various TDI data combinations for TianQin. The optical paths are evaluated based on the numerical 
orbits of TianQin's satellites that have been optimized 
to meet the orbital stability requirements imposed by the long range space laser interferometry. 
Four types of the first-generation TDI data combinations show time differences of 
$\sim 10^{-8}$~s which makes them competent for GW signal detection for the frequencies 
$\lesssim 10^{-3}$~Hz and  $\gtrsim 10^{-1}$~Hz given the stabilization of the laser frequency noise of  
$10~\mathrm{Hz}/\sqrt{\mathrm{Hz}}$ in concerned frequency range. 
With an ample margin, the second-generation TDI 
data combinations with time differences of $\sim 10^{-12}$~s 
are warranted to reduce the laser frequency noise well below the 
secondary noises.

The rest of the paper is organized as follows. In section \ref{sec:orbit}, 
we discuss the orbit optimization for TianQin's satellites. 
The resulting numerical orbits interpolated by Chebyshev polynomials 
are subsequently used in the simulations of the time differences of the symmetric interference paths
for various first- and second-generation TDI data combinations 
in section \ref{sec:sim1} and \ref{sec:sim2}, respectively. 
The paper is concluded in section \ref{sec:sum}.

\section{Constellation optimization for TianQin }\label{sec:orbit}

The orbit of each satellite is primarily determined by the monopole gravitational field
of the Earth. Besides, the perturbation from the multipole terms and relativistic
post-Newtonian correction of the Earth gravitational field, the monople gravitational field of
the Moon, the Sun, the major planets in the solar system, Pluto and large asteroids
will also contribute. Dispersion from the Earth atmosphere and solar radiation are
ignored due to the implementation of the drag-free control for the satellite platform.
The minor eccentricity of the nominal Keplerian orbit along with time-dependent perturbation forces
will induce variation of the armlengths, and flexing and breathing of the triangular constellation.
On the other hand, the high precision space laser interferometry imposes requirements on
the stability of the triangular constellation
\cite{2016CQGra..33c5010L, 2019IJMPD..2850121Y}:
(a) the armlength variation less than $0.5\% \times (\sqrt{3} \times 10^5)~\mathrm{km}$; 
(b) the breathing angle (subtended by two arms) variation less than $0.1^{\circ}$ during the first two years and
less than $0.2^{\circ}$ during the five years mission lifetime;
(c) the relative range rate (Doppler velocity) less than $5~\mathrm{m/s}$ during the first two years and
less than $10~\mathrm{m/s}$ in five years.

\subsection{Calculation of the satellite orbits}\label{subsec:calorb}

\begin{figure}[!h]
\centering
\includegraphics[width=0.6\textwidth]{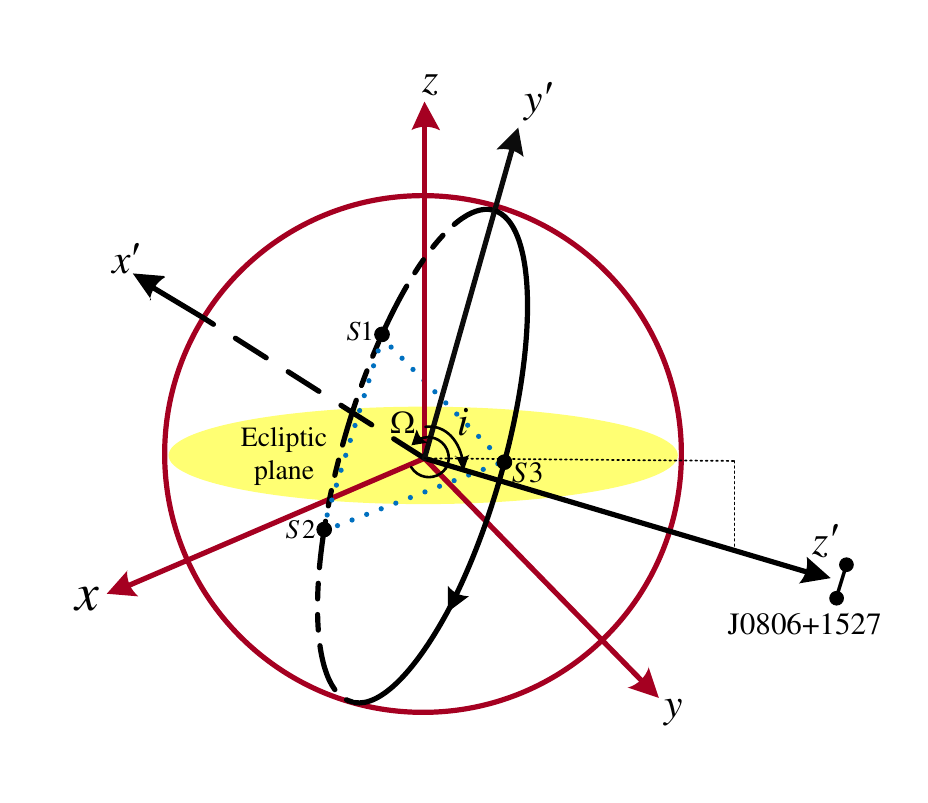}
\caption{\label{fig:1} $(x, y, z)$ represents the geocentric ecliptic coordinate system, 
$x$ axis points toward the mean equinox of J2000.  
$(x', y', z')$ represents the orbit coordinate
system of the detector, $x'$ axis points toward the ascending node of the satellites and $z'$ axis
points toward the reference source RX J0806+1527.  $\Omega$ is the angle between
$x$ and $x'$, $i$ is the inclination of the detector plane.}
 \end{figure}

In this work, the geocentric ecliptic coordinate system ($x,y,z$) shown in Fig.~\ref{fig:1}
has been adopted in the calculation and optimization of the orbits for TianQin's satellites. 
This choice is different from \cite{Hu2018} in which, for the sake of calculating 
the antenna response of TianQin to GWs, 
the heliocentric ecliptic coordinate system is used. 
The gravitational field of the Earth is described in the Earth-fixed reference coordinate system WGS84 \cite{1997isbb.book.....E}.

From the six initial Keplerian elements $\sigma_{0}=(a_{k}, e_{k}, i_{k}, \Omega, \omega, M_{k})$
of the $k$-th ($k \in \lbrace 1,2,3 \rbrace$) satellite,
we can obtain the initial Cartesian coordinates in the geocentric
ecliptic coordinate system as follows \cite{2005ormo.book.....R}: 
\begin{equation}\label{eq:cartesian}
\begin{aligned}
\begin{array}{l}
x_{k}=l_{1}a_{k}(\cos E_{k}-e_{k})+l_{2}a_{k}\sqrt{1-{e_{k}}^{2}}\sin E_{k}  \,, \\
y_{k}=m_{1}a_{k}(\cos E_{k}-e_{k})+m_{2}a_{k}\sqrt{1-{e_{k}}^{2}}\sin E_{k} \,,  \\
z_{k}=n_{1}a_{k}(\cos E_{k}-e_{k})+n_{2}a_{k}\sqrt{1-{e_{k}}^{2}}\sin E_{k}  \,.
\end{array}
\end{aligned}
\end{equation}
Here $a_{k} \approx 1.0 \times10^{5}~\mathrm{km}$ is the semi-major axis,  
and $e_{k} \approx 0$ is the orbit eccentricity.
The eccentric anomaly $E_{k}$ can be obtained by the Newton's
iteration method \cite{2002isbn.book.....E}:
\begin{equation}\label{eq:eccanom}
E_{n+1,k}=E_{n,k}+\frac{M_{k}-E_{n,k}+e_{k}\sin E_{n,k}}{1-e_{k}\cos E_{n,k}} \,,
\, ~(n=0,1,2, \cdots)  \,,
\end{equation}
where $n$ is the index of the iteration and $E_{0,k} = M_k$.
We set the mean anomaly $M_{k}=\varpi_k (t-t_{0})+120^{\circ}\times(k-1)+60^{\circ}$
in order to arrange the three satellites into a nearly equilateral triangle constellation.
$\varpi_k = \sqrt{GM_{\oplus}/a_{k}^{3}}$ represents the angular velocity of the satellite, 
where $M_{\oplus}$ is the Earth mass. 
The surrogate variables $(l_{1,2},m_{1,2},n_{1,2})$ in Eq.~\ref{eq:cartesian} are defined as follows:
\begin{equation}
\begin{aligned}
\begin{array}{l}
l_{1}=\cos \Omega \cos \omega-\sin \Omega \sin \omega \cos i_{k} \,, \\
m_{1}=\sin \Omega \cos \omega+\cos \Omega \sin \omega \cos i_{k}  \,,\\
n_{1}=\sin\omega\sin i_{k}  \,; \\
l_{2}=-\cos \Omega \sin \omega-\sin \Omega \cos \omega \cos i_{k}  \,, \\
m_{2}=-\sin \Omega \sin\omega+\cos \Omega\cos \omega \cos i_{k} \,, \\
n_{2}=\cos\omega\sin i_{k} \,.
\end{array}
\end{aligned}
\end{equation}
Here we set the argument of periapsis (the angle between ascending node 
and perigee) $\omega=0$.
The three satellites form a detector plane (subtended by $x'$ and $y'$ axes in Fig.~\ref{fig:1}),
the normal of which ($z'$ axis) points toward the reference source RX J0806+1527. 
It thus indicates that the inclination of the detector plane $i_{k}=94.7^{\circ}$
and the longitude of the ascending node (the angle between $x$ axis and 
$x'$ axis) $\Omega=210.5^{\circ}$.

Finally, through coordinate transformation, we obtain the expression for the initial positions 
of the satellites in geocentric equatorial coordinates. Here we choose obliquity of
the ecliptic $\varepsilon=23^{\circ}26^{''}$.

The subsequent evolution of the satellites' orbits, after setting up the initial condition,
is determined by the dynamic equation \cite{1997isbb.book.....E}
(for clarity we ignore the satellite index $k$ hereafter), 
\begin{equation}\label{eq:acc3body}
\mathbf{\ddot{r}} =\mathbf{\ddot{r}}_{\rm{NB}}+\mathbf{\ddot{r}}_{\rm NS} +\mathbf{\ddot{r}}_{\rm{PN}}  \,,
\end{equation}
where $\mathbf{r}$ represents the position vector of the satellite relative to
the geocenter.  The Newtonian body term is
\begin{equation}\label{eq:forces}
\mathbf{\ddot{r}}_{\rm{NB}}=-GM_{\oplus}\mathbf{r}/r^{3}-\sum_{i=1}^{10}Gm_{i}(\mathbf{\Delta}_{i}/\Delta_{i}^{3}+\mathbf{r}_{i}/r^{3}_{i})  \,,~  (i=1,2,3, \cdots, 10)  \,.
\end{equation}
Here $G$ is the gravitational constant. 
$m_{i}$, in the ascending order of $i$, denotes the masses 
of the seven major planets, Pluto, the Sun and the Moon.
$\mathbf{r}_{i}$ and $\mathbf{\Delta}_{i} =\mathbf{r} - \mathbf{r}_{i} $ are the position vectors
of the $i$-th perturbing body relative to the geocenter and the satellite, respectively.
$r_{i} = |\mathbf{r}_{i}|$ and  $\Delta_{i} = |\mathbf{\Delta}_{i}|$.
The positions of the planets, the Sun, the Moon are extracted from 
the ephemeris DE421 \cite{2009IPNPR.178C...1F}. 
$\mathbf{\ddot{r}}_{\rm NS}$ includes the sectorial and tesseral
harmonic terms of the Earth's non-spherical gravitational perturbation, respectively.
The first-order post-Newtonian correction 
term \cite{1993tegp.book.....W,1968PhRv..169.1017N,1972ApJ...177..757W} is  
\begin{equation}\label{eq:pn}
\mathbf{\ddot{r}}_{\rm{PN}}=\frac{GM_{\oplus}}{c^{2}r^{2}}\left[ \left( 4GM_{\oplus}/r-v^{2} \right) \left(\mathbf{r}/r\right)+4\left(\mathbf{r} \cdot \dot{\mathbf{r}}\right) \left(\dot{\mathbf{r}}/r\right) \right]   \,,
\end{equation}
where $v= |\dot{\mathbf{r}}|$, and $c$ is the speed of light.

Once set the orbit elements at an initial epoch, for example
$t_{0}=$ MJD 64104.5 (May 22 2034 12:00:00 TDB),
we can numerically integrate Eq.~\ref{eq:acc3body} by Runge-Kutta 7(8) method to obtain $\mathbf{r}$
and $\mathbf{\dot{r}}$ at subsequent
time stamps $t_i$ ($i=1,2, \cdots, N$), where $N=T/\Delta t$ and $T=5~\mathrm{yr}$.
The integration time stepsize $\Delta t=1~\mathrm{hr}$, which is chosen such that the
total computational error (combining the accumulated round-off error from
arithmetic operations of double-precision float number and
the truncation error of Runge-Kutta algorithm) approaches its minimum.

\subsection{The constellation optimization}\label{subsec:opt}

In order to fulfill the aforementioned requirements on orbit stability,
we try to find a set of initial orbit elements in the vicinity of the fiducial ones such that
during the mission lifetime the armlength variation measured by the sum of
squared length differences of the three arms reaches a global minimum in the
parameter space. 
It has been proved effective in practice to only search in the nine dimensional parameter space spanned by
$\bar{\sigma}_{0} = \left(a_{k}, i_{k},e_{k}\right)$, $k \in \lbrace 1,2,3 \rbrace$ \cite{Hu2015}. 
Thus, the constellation optimization problem can be formulated as follows,
\begin{equation}\label{eq:optele}
\bar{\sigma}_{0}^{\ast} = \argmin_{\bar{\sigma}_{0}\in\mathbb{D}} \mathcal{O}(\bar{\sigma}_{0}) = \argmin_{\bar{\sigma}_{0}\in\mathbb{D}} \sum_{k=1}^{3}\sum_{i=1}^{N}(L_{k,i}-L_{k,0})^{2} \,,
\end{equation}
where $L_{k,i}$ represents the length of the arm opposite to the $k$-th satellite
at time $t_{i}$, and $L_{k,0}$ is the armlength at the initial epoch.
$\mathbb{D}$ is the \textit{search space} of the optimization 
with the ranges: $a_{k}\in[0.999, 1.001]\times10^{5}$~km, $e_{k}\in[0, 1\times10^{-3}]$, $i_{k}\in[93.2^{\circ}, 96.2^{\circ}]$.

The \textit{objective} function $\mathcal{O}(\bar{\sigma}_{0})$ may have a highly
multi-modal landscape owning a forest of local minima. Therefore
any deterministic local optimizer would locate a suboptimal solution. On the other hand, a direct grid search
will be computationally prohibitive due to the high dimension ($\dim(\mathbb{D})=9$).
Instead, the optimization methods including some randomness can be used to effectively pinpoint the
global minimizer in $\mathcal{O}(\bar{\sigma}_{0})$. 
In this work, we use one of the stochastic optimization methods based
on emulating biological groups, namely the particle swarm optimization (PSO) \cite{eberhart1995new}.
PSO has been applied in many fields \cite{mohanty2018swarm},
including gravitational wave data analysis for detecting and estimating
compact binary coalescence signals in a network of ground-based
laser interferometers \cite{2017PhRvD..95l4030W,2018PhRvD..98d4029N}
and continuous waves in pulsar timing arrays \cite{2014ApJ...795...96W, 2015ApJ...815..125W}.

\begin{figure}[!htp]
\centering
    \subfigure[ ] {
    \begin{minipage}[t]{0.5\linewidth}
    \centering 
    \includegraphics[height=6.8cm,width=8.2cm]{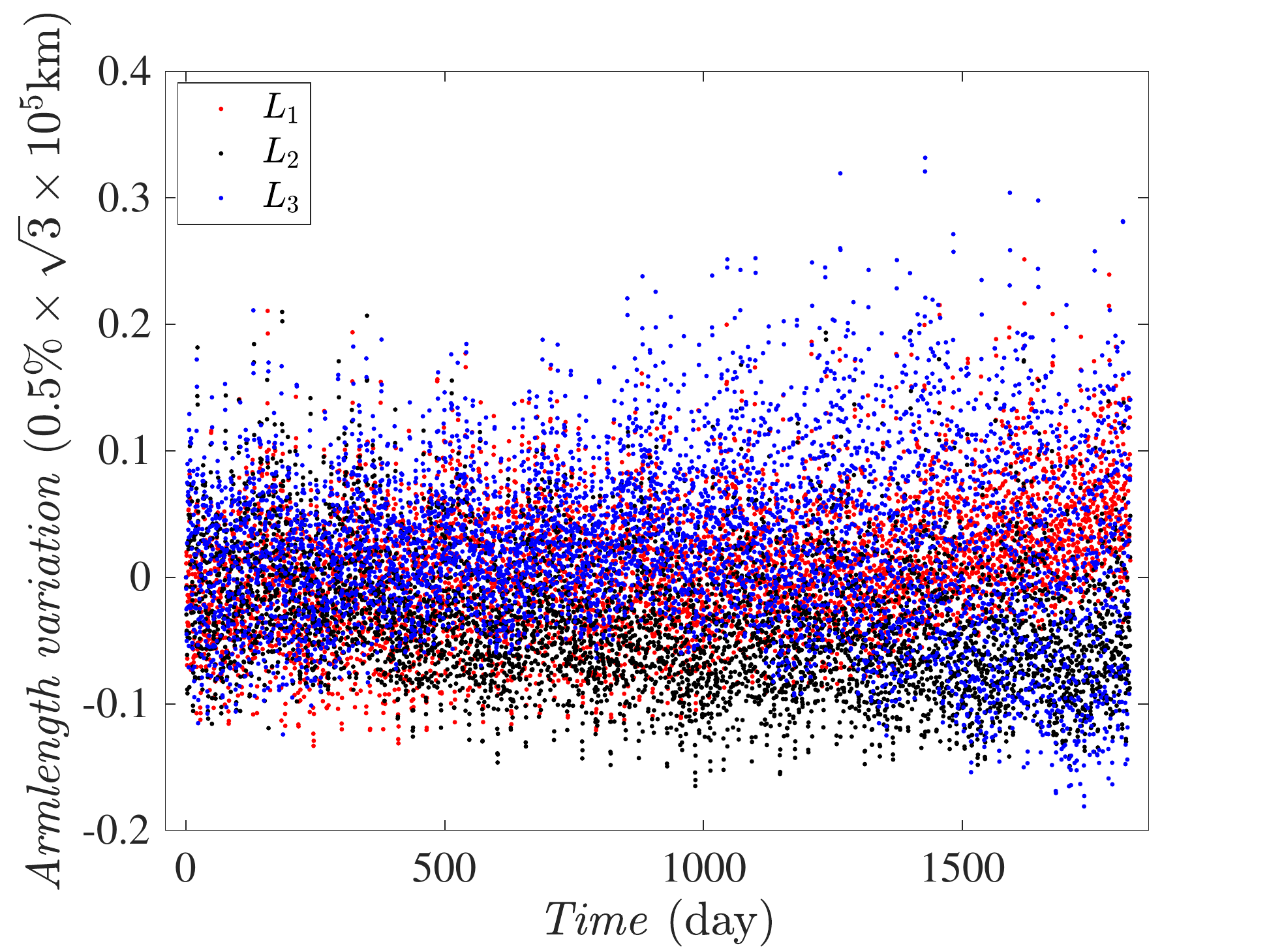} 
    \end{minipage} %
    }%
    \subfigure[ ]{
    \begin{minipage}[t]{0.5\linewidth}
    \centering 
    \includegraphics[height=6.8cm,width=8.2cm]{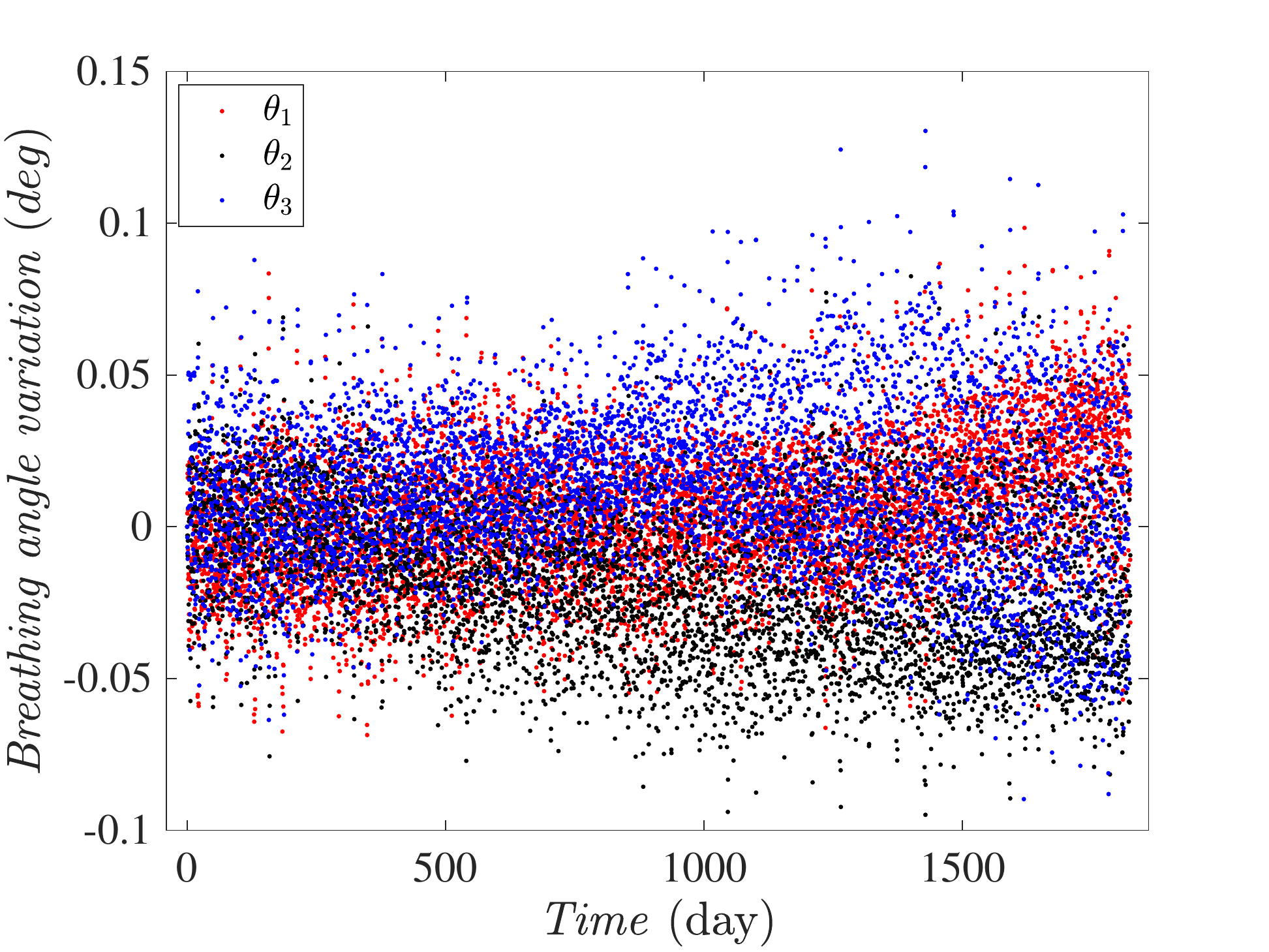} 
    \end{minipage}%
    } %
    
    \subfigure[ ] {
    \begin{minipage}[t]{0.5\linewidth}
    \centering 
    \includegraphics[height=6.8cm,width=8.2cm]{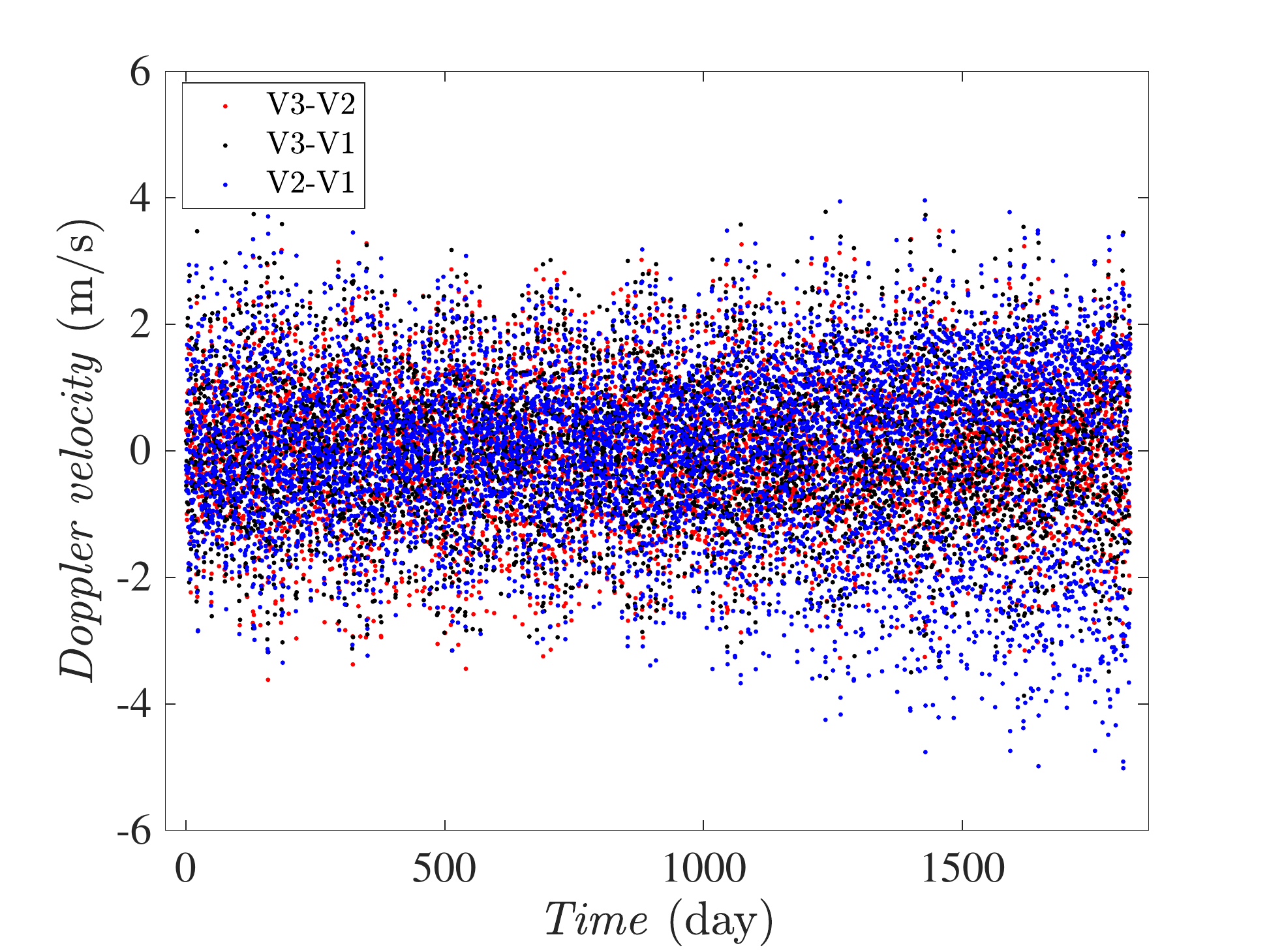} 
    \end{minipage} %
    }%
    \subfigure[ ] {
    \begin{minipage}[t]{0.5\linewidth}
    \centering 
    \includegraphics[height=6.8cm,width=8.2cm]{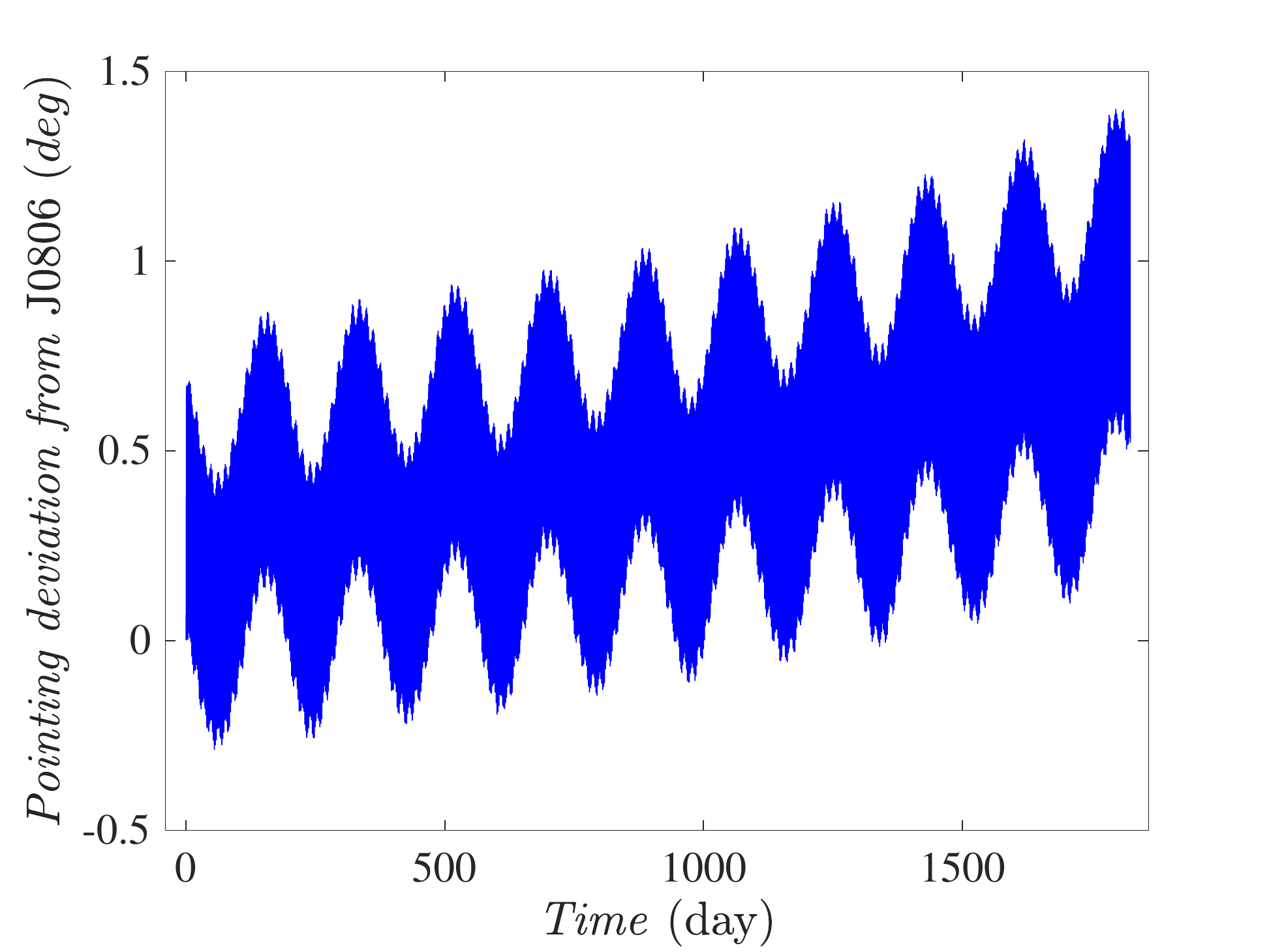} 
    \end{minipage}%
    } %
\centering  
    \caption{\label{fig:orbit} $(a)$ The variations of armlengths; $(b)$ The variations of breathing angles; $(c)$ Doppler velocities; $(d)$ Pointing deviation from J0806+1527 during 5-yr mission lifetime.}
    \end{figure}

The resulting $\bar{\sigma}_{0}^{\ast}$ for each satellite is listed in Table \ref{tab:tab1}.
The variations of the armlengths, the variations of breathing angles, Doppler velocities 
and pointing deviation (from the direction of RX J0806+1527) 
of the satellite constellation during 5-yr mission lifetime are shown in Fig.~\ref{fig:orbit}. 
We can see that the optimized orbits satisfy the aforementioned requirements on
the constellation stability with some margins. 
Note that, although the constellation optimization method is different from the three-step scheme
adopted in \cite{2019IJMPD..2850121Y},
the results, in terms of the initial elements and performance
of constellation optimization is virtually consistent (see Fig.~1 in \cite{2019IJMPD..2850121Y}).

\begin{table*}
\caption{\label{tab:tab1} The values of the elements of the optimized orbit $\bar{\sigma}_{0}^{\ast}$ 
for the three satellites at the initial epoch $t_{0}=$ MJD 64104.5.}
\centering
\begin{tabular}{cccc}
  \hline
  \hline
   &~Semi-major axis~ & ~Inclination~ & ~Eccentricity~ \\
   & $a$~(km) & $ i~(^{\circ})$ & $e$ \\
   \hline
  S1&99995.0717&94.7473&
  $2.8483\times10^{-4}$ \\
  \hline
  S2 &100010.8914&95.7094&
  $0$ \\
  \hline
  S3 &99992.5623&94.6469&
  $2.0876\times10^{-4}$ \\
  \hline
\end{tabular}
\end{table*}

\subsection{Interpolation of the orbit}\label{subsec:interp}

The orbit coordinates of the satellites can be represented as  finite sequences 
with sampling interval of $\Delta t$.
As it will become more clear in Sec.~\ref{sec:sim}, the orbit positions and velocities at arbitrary time
points within $\left( t_i, t_{i+1} \right)$ are needed in generating the TDI data combinations.
In this case, we use Chebyshev polynomials to approximate the orbit of each satellite
at $t\in ( t_i, t_{i+1})$. 
This method is stable and has been widely used in high precision ephemeris
interpolation, such as the DE series ephemerides developed by Jet Propulsion Laboratory (JPL) \cite{1989CeMec..45..305N}.
Here, Chebyshev polynomials up to 15 orders are adopted and the sampling interval 
of the fitted data is 0.5 day with interpolation precision of $\sim 10^{-10}$ km. 
The weight of position and velocity is $1:0.4$, which turns out to be the optimal choice
to calculate the positions of the planets in the solar system \cite{1989CeMec..45..305N}. The function approximation algorithm finds the best-fit coefficients of
Chebyshev polynomials by minimizing the variance of the residuals 
between the data and the model \cite{2007nras.book.....P}.

\section{Simulation of time differences for TDI data combinations} \label{sec:sim} 

\begin{figure}[!h]
\centering
\includegraphics[width=0.4\textwidth]{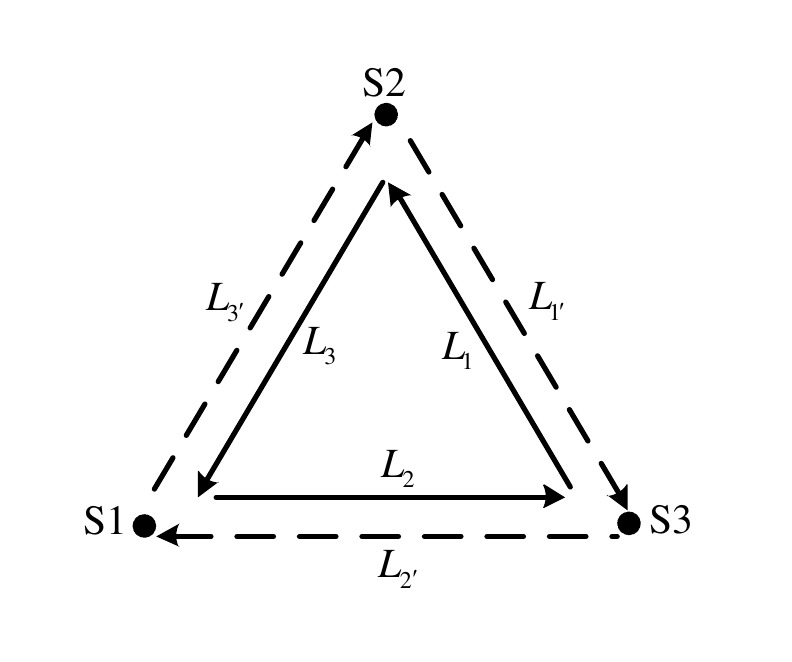}
\caption{\label{fig:3} The labels of the satellites and the interference arms.}
\end{figure}

In this work, we assume that the lasers on the two optical benches 
housed in a satellite are locked in phase, so we only 
consider the inter-satellite fractional frequency measurements `$y$' which account for 
the cancellation of the laser frequency noise. 
Following \cite{2005LRR.....8....4T}, the naming convention of the interference arms is shown in Fig.~\ref{fig:3}.

\subsection{Time differences for the first-generation TDI}\label{sec:sim1}

\begin{figure}[!h]
\centering
\includegraphics[width=0.5\textwidth]{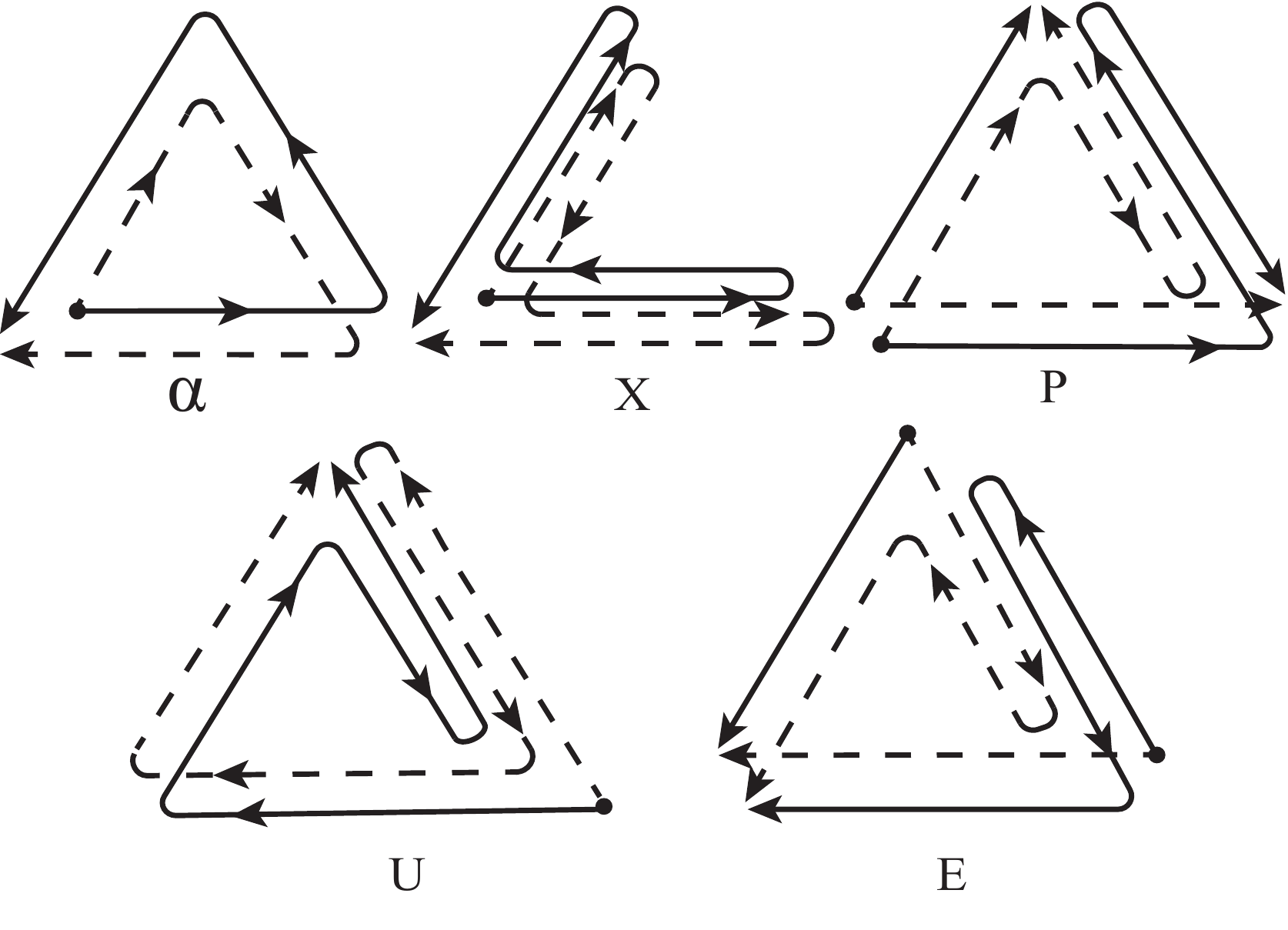}
\caption{\label{fig:4} Schematic diagram of the first-generation TDI combinations.}
\end{figure}

\begin{figure}[!htp]
\centering
\includegraphics[height=6.8cm,width=8.5cm]{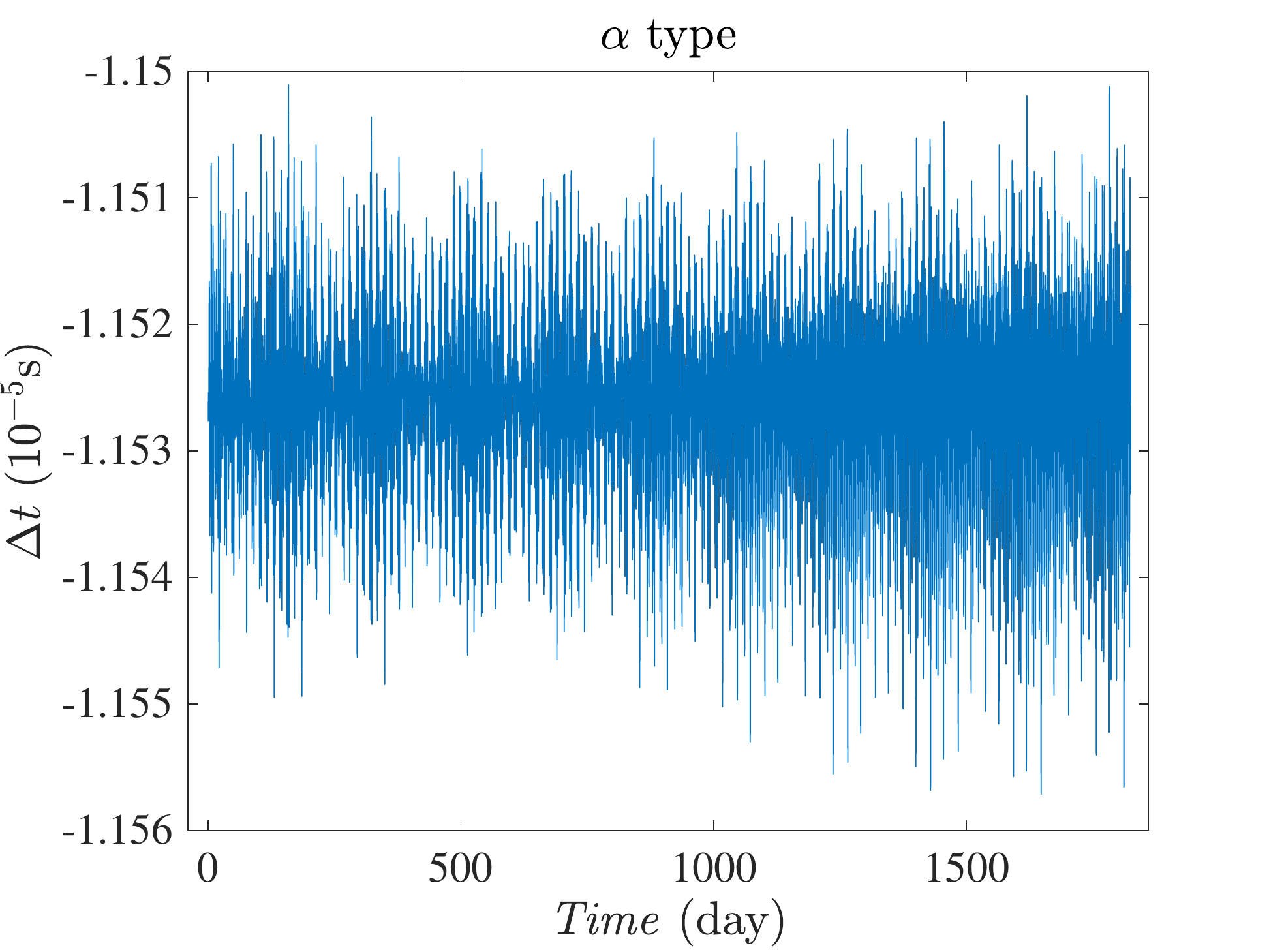}
\caption{\label{fig:alpha} Simulation of the time difference for the first-generation TDI $\alpha$ combination. }
\end{figure}

The first-generation TDI has 15 interference data combinations, namely $(\alpha,\beta,\gamma)$, $(X,Y,Z)$,  
$(U,V,W)$, $(E,F,G)$, and $(P ,Q,R)$. 
Five representative combinations are shown in Fig.~\ref{fig:4}. 
The other subtypes are different in the starting satellite of light paths and their combinations can be obtained by 
cyclic permutation of the indices for satellites: $1\rightarrow2\rightarrow3\rightarrow1$.  
As an example, the Doppler data for the Sagnac-type $\alpha$ combination is as follows:
\begin{equation} \label{eq:fTDIal}
\alpha=y_{23,13}+y_{12,3}+y_{31}-y_{21}-y_{13,2'}-y_{32,1'2'}  \,,
\end{equation}
where `,' marks the time delay of the laser beam traveling along an arm. 
$y_{23,13}=y_{23}(t-L_{1}-L_{3})$, which represents the time-delayed 
fractional frequency fluctuation time series measured at reception by S3 with transmission from S1 along arm $2$. 
Here and hereafter, we set $c=1$. 
As in Fig.~\ref{fig:4}, the clockwise (1-2-3-1) and the counter clockwise (1-3-2-1) light paths 
interfere at S1 at time $t$. While, their initial times of emission at S1 are 
$t-L_{3'}-L_{1'}-L_{2'}$ and $t-L_{2}-L_{1}-L_{3}$, respectively.
Inserting the fractional frequency fluctuation $C_{i}(t)$ of the laser on S$i$ into Eq.~\ref{eq:fTDIal}, we obtain:
 \begin{equation} \label{eq:fTDIale}
 \begin{aligned}
 \delta C_{\alpha}(t)=&[C_{1}(t-L_{2}-L_{1}-L_{3})-C_{3}(t-L_{1}-L_{3})]+[C_{3}(t-L_{1}-L_{3})-C_{2}(t-L_{3})] \\
 &+[C_{2}(t-L_{3})-C_{1}(t))]-[C_{3}(t-L_{2'})-C_{1}(t)] \\
 &-[C_{2}(t-L_{1'}-L_{2'})-C_{3}(t-L_{2'})]\\
 &-[C_{1}(t-L_{3'}-L_{1'}-L_{2'})-C_{2}(t-L_{1'}-L_{2'})]\\
 &=C_{1}(t-L_{2}-L_{1}-L_{3})-C_{1}(t-L_{3'}-L_{1'}-L_{2'}) \,.
 \end{aligned}
 \end{equation}
The first-generation TDI assumes a fixed detector constellation in space and  
the armlengths are unequal but constant, 
thus Eq.~\ref{eq:fTDIale} can be canceled out exactly. 
However, in reality, the armlengths are time varying due to the actual rotating and 
flexing of the constellation. 
To the first order of $L_{l}(t)$, Eq.~\ref{eq:fTDIale} can be expanded as: 
 \begin{equation} \label{eq:fTDIale2}
 \begin{aligned}
\delta C_{\alpha}(t) &\simeq \dot{ C}_{1}(t)[-L_{2}+\dot{L}_{2}(L_{1}+L_{3})-L_{1}+\dot{L}_{1}L_{3}-L_{3}\\
 &+L_{3'}-\dot{L}_{3'}(L_{1'}+L_{2'})+L_{1'}-\dot{L}_{1'}(L_{2'})+L_{2'}]\\
 \end{aligned}
 \end{equation}
where both $L_{l}$ and $\dot{L}_{l}$ are evaluated at time $t$. 
When only the zero order of $ L_{l}(t)$ is concerned, as in the first-generation TDI, 
Eq.~\ref{eq:fTDIale2} vanishes. 
The data combinations that have this nature are called \textit{L-closed} \cite{2005PhRvD..72d2003V}. 
Therefore, the level of laser frequency noise cancellation is determined by how well 
we can synthesize equal-length virtual paths, such as the two paths for 
the $\alpha$ combination, in the construction of virtual interferometers. 
In other words, from Eq.~\ref{eq:fTDIale}, we can see that the magnitude of the 
residual laser noise can be measured by the time difference of the two interference paths, 
which, for the $\alpha$ combination, is 
 \begin{equation}
 \begin{aligned}
 \Delta t_{\alpha}=L_{2,13}+L_{1,3}+L_{3}-L_{2'}-L_{1',2'}-L_{3',1'2'} \,. 
 \end{aligned}
 \end{equation}
Here, $L_{2,13}=L_{2}(t-L_{1}(t)-L_{3}(t))$ is the length of $L_{2}$ at time $t-L_{1}(t)-L_{3}(t)$. 
Through the armlengths obtained in Sec.~\ref{subsec:opt}, we can calculate 
the time differences of interference paths, which can be used to analyze the 
level of the laser frequency noise cancellation. Fig.~\ref{fig:alpha} presents the result for the $\alpha$ 
combination in the five years mission lifetime. 
We can see the net time difference of the two opposite paths induced by the Sagnac effect 
$\Delta t_{\rm Sag} = 4 \vec{\Omega} \cdot \vec{A}$ \cite{2003CQGra..20.4851C,2004PhRvD..69b2001S}. 
With the area $|\vec{A}| = \sqrt{3}L^{2}/4$ ($L \simeq \sqrt{3} \times 10^{5} ~\mathrm{km}$) 
and the angular velocity of the constellation $|\vec{\Omega}|=1/3.65 ~\mathrm{cycle/day}$ for TianQin, 
$\Delta t_{\rm Sag} \simeq 1.15 \times 10^{-5}~\mathrm{s}$. 
Therefore, the result of this data combination is about three orders of magnitude worse than the other first-generation data combinations below. 

\begin{figure}[!htp]
\centering
    \subfigure[ ] {
    \begin{minipage}[t]{0.5\linewidth}
    \centering 
    \includegraphics[height=6.8cm,width=8.2cm]{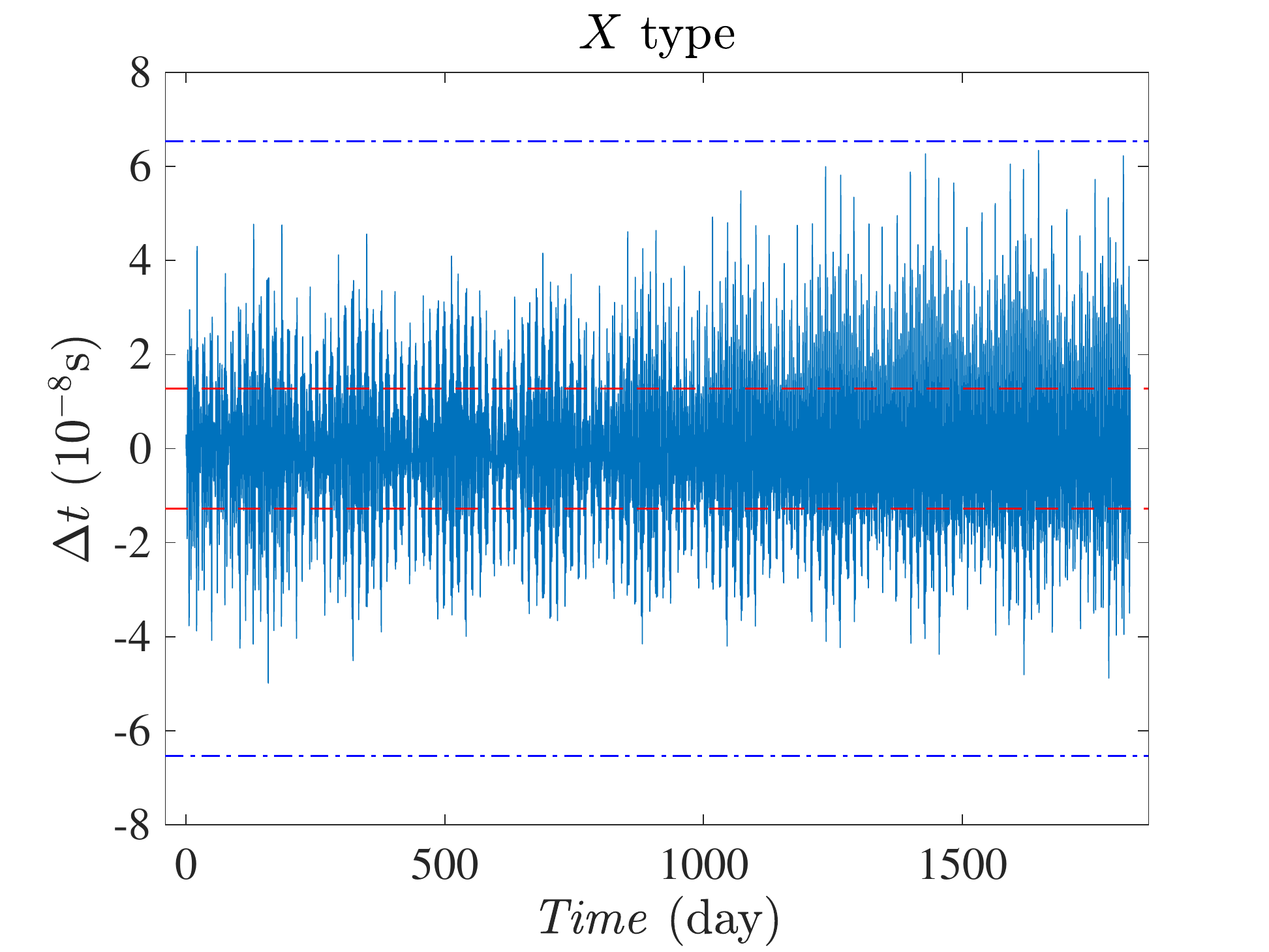} 
    \end{minipage} %
    }%
    \subfigure[ ]{
    \begin{minipage}[t]{0.5\linewidth}
    \centering 
    \includegraphics[height=6.8cm,width=8.2cm]{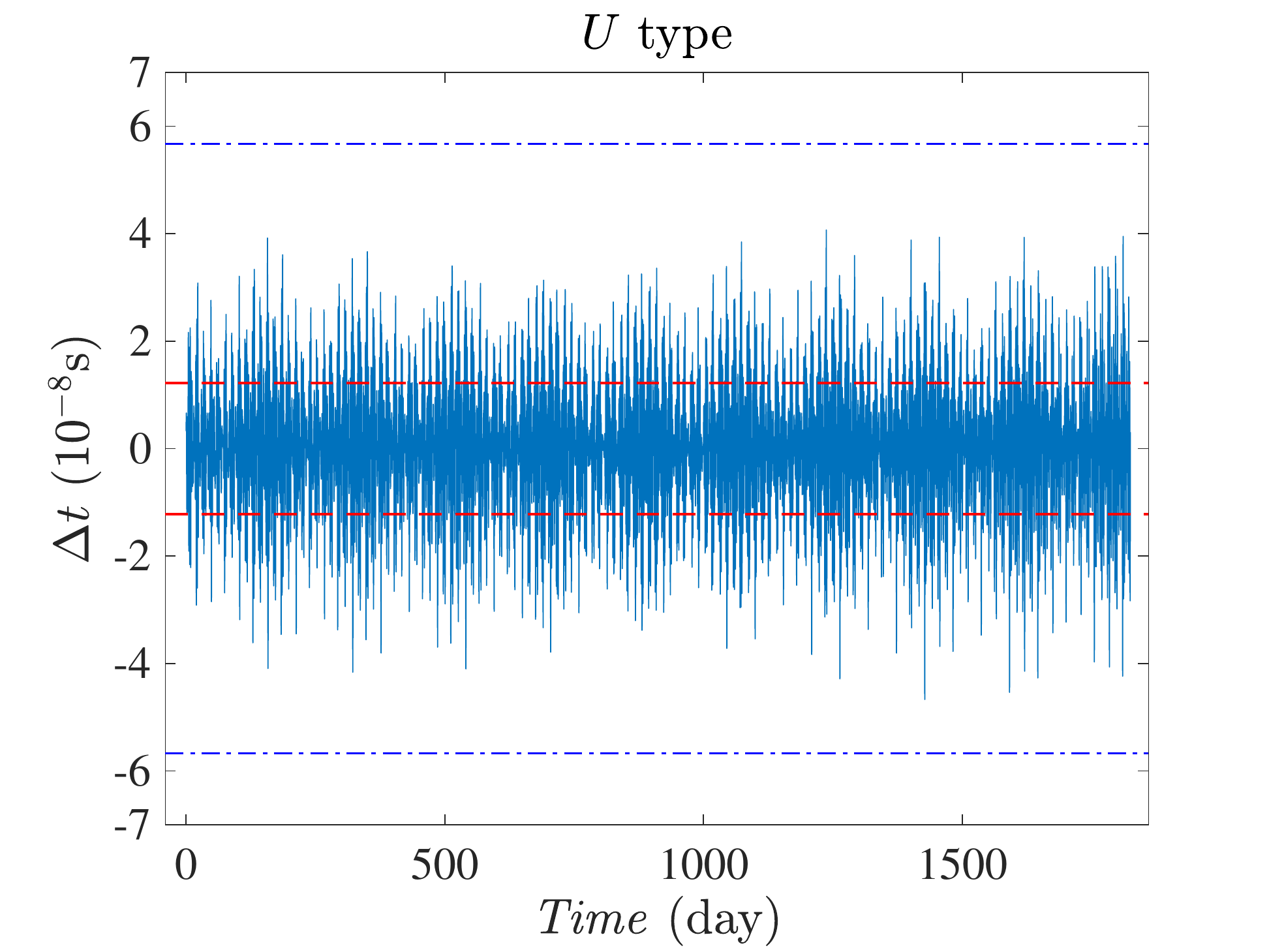} 
    \end{minipage}%
    } %
    
    \subfigure[ ] {
    \begin{minipage}[t]{0.5\linewidth}
    \centering 
    \includegraphics[height=6.8cm,width=8.2cm]{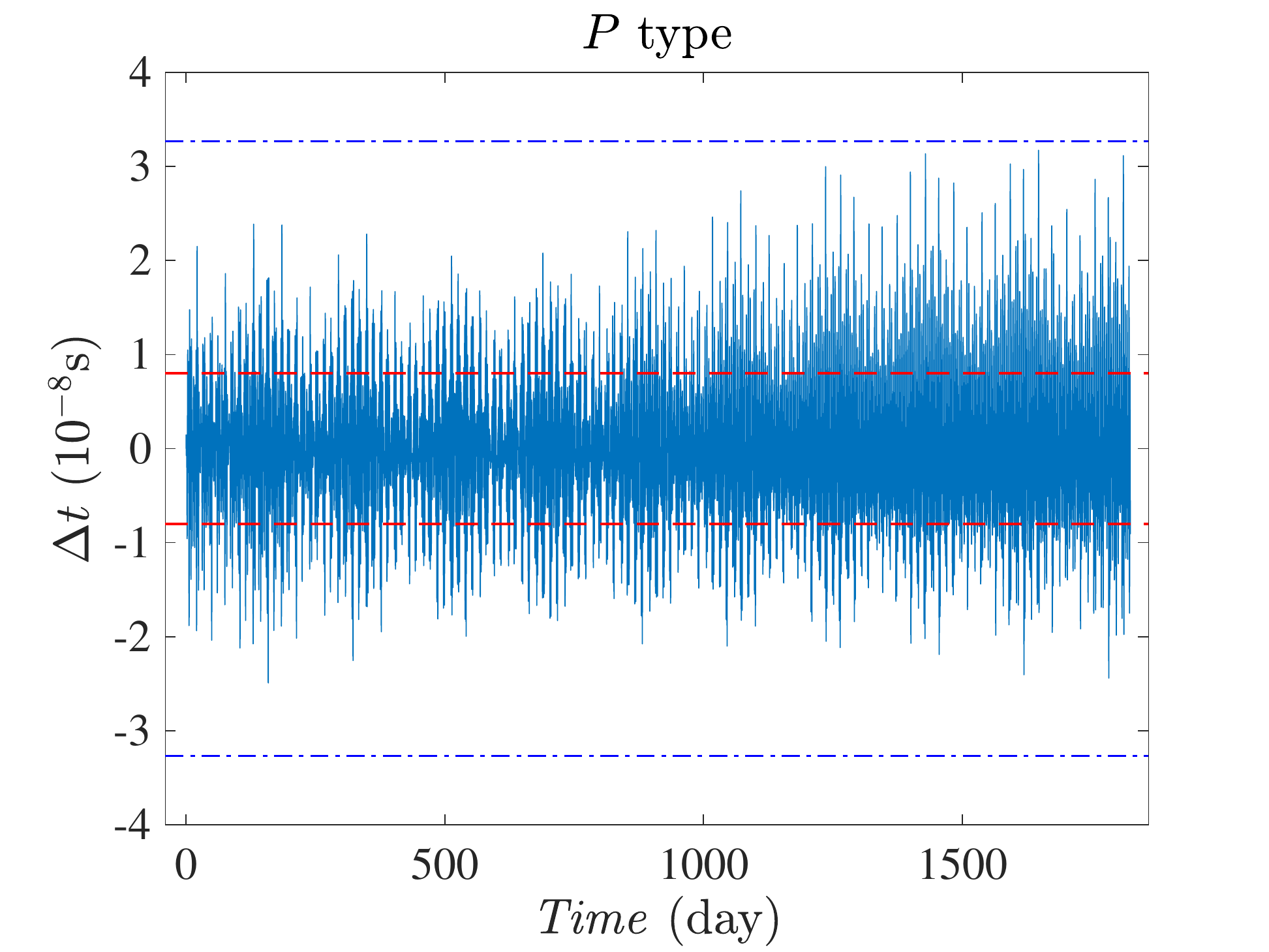} 
    \end{minipage} %
    }%
    \subfigure[ ] {
    \begin{minipage}[t]{0.5\linewidth}
    \centering 
    \includegraphics[height=6.8cm,width=8.2cm]{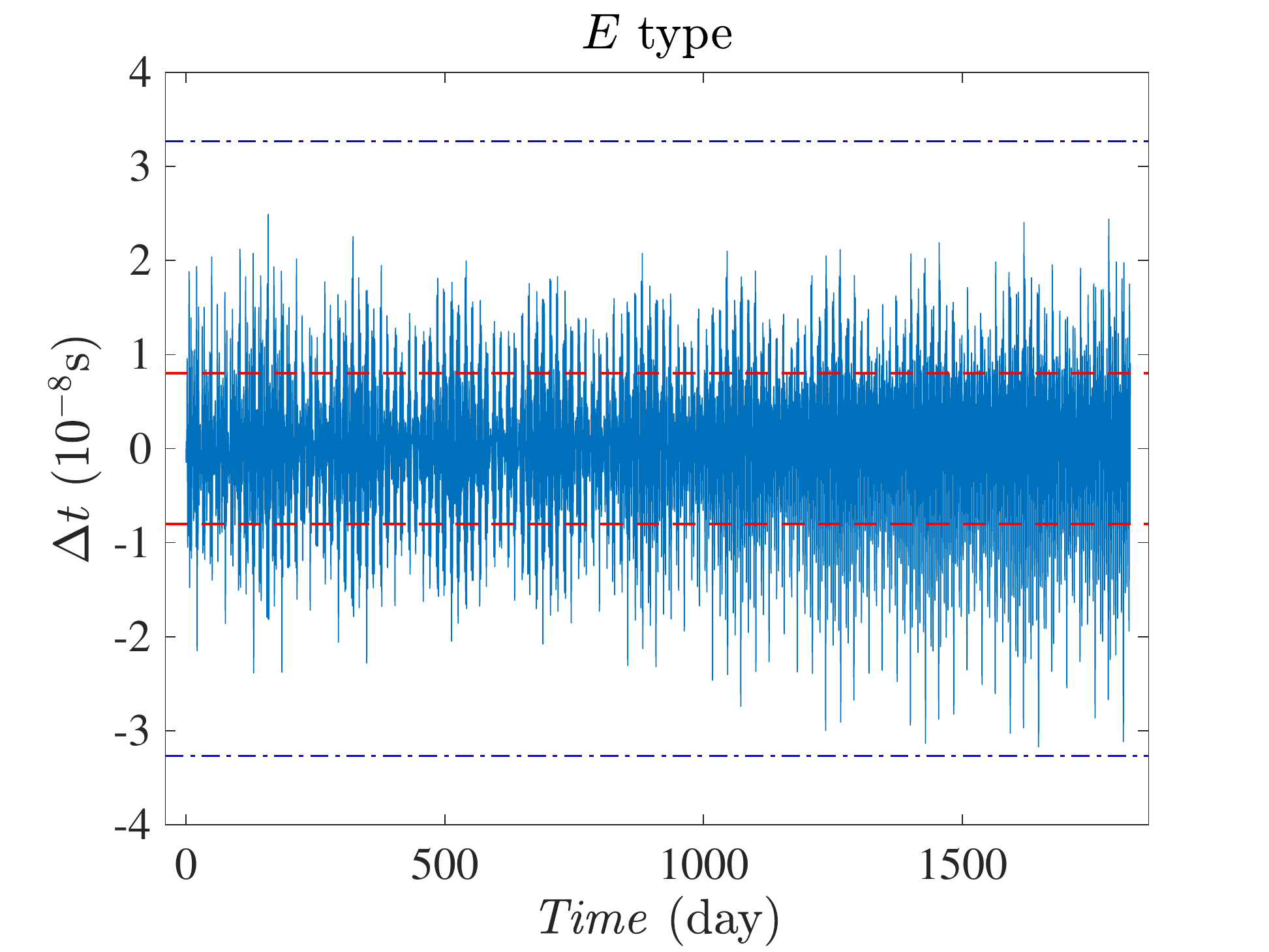} 
    \end{minipage}%
    } %
\centering  
    \caption{\label{fig:firstTDI} Simulation of the time differences for the first-generation TDI $X$, $U$, $P$ and $E$  combinations (blue solid line). In each panel, the red dash lines represent the lowest 
required time differences for each combinations at $f\approx 10^{-2.2}$~Hz (c.f. right panel of Fig. \ref{fig:secondary_noise}). 
For comparison, the blue dash-dot lines represent the ones at $f = 10^{-3}$~Hz. }
    \end{figure}

Following the same procedure, the time differences for the other types can be written as: 
 \begin{equation}
 \begin{aligned}
 &\Delta t_{X}=L_{3',322'}+L_{3,22'}+L_{2,2'}+L_{2'}-L_{3}-L_{3',3}-L_{2',3'3}-L_{2,2'3'3} \,, \\
 &\Delta t_{U}=L_{2',3'1'1}+L_{3',1'1}+L_{1',1}+L_{1}-L_{3'}-L_{2',3'}-L_{1',2'3'}-L_{1,1'2'3'} \,,\\
 &\Delta t_{P}=L_{3',1'12}+L_{1',12}+L_{1,2}-L_{3',2}+L_{2,3'}-L_{1',3'}-L_{1,1'3'}-L_{2,11'3'} \,, \\
 &\Delta t_{E}=L_{1',13}+L_{1,3}+L_{3}-L_{2'}-L_{1',2'}-L_{1,1'2'}+L_{2',1'1}-L_{3,11'} \,.
 \end{aligned}
 \end{equation}
The corresponding results are shown in Fig.~\ref{fig:firstTDI}. The time differences for the first-generation TDI 
$X$, $U$, $P$ and $E$ data combinations are $\sim 10^{-8}$ s. 

The TDI data combinations can be considered effective when, taking the $\alpha$ combination as 
an example, $|\delta \tilde{C}_{\alpha}(f)|^{2} \leq S_{\alpha}(f)$. Here, 
\begin{equation}\label{eq:alpha}
|\delta \tilde{C}_{\alpha}(f)|^{2} = 4 \pi^{2} f^{2} |\tilde{C}_{1}|^{2} \delta t_{\alpha}^{2} \,, 
\end{equation}
which can be obtained by using the derivative theorem of Fourier transform for Eq. \ref{eq:fTDIale} \cite{2005PhRvD..71b2001V}. 
$\delta t_{\alpha} \equiv \sup \left\lbrace |\Delta t_{\alpha}(t)| \right\rbrace$. 
$S_{\alpha}(f) =[4 \sin^{2}(3 \pi f L) + 24 \sin^{2}(\pi f L) ]S^{\rm accel}_{y} + 6S^{\rm opt}_{y}$ 
is the secondary noise power spectral density (PSD) for the $\alpha$ combination \cite{1999ApJ...527..814A},  
where $S^{\rm accel}_{y} =2.8\times10^{-49}(f/1~\mathrm{Hz})^{-2}~\mathrm{Hz}^{-1}$ 
and $S^{\rm opt}_{y}= 4.4 \times 10^{-40} (f/1~\mathrm{Hz})^{2}~\mathrm{Hz}^{-1}$ 
are the PSDs of the acceleration noise and the optical path noise \cite{1999ApJ...527..814A} 
assuming the one-sided amplitude spectra of the acceleration noise and the optical path noise of a single link 
are $1\times10^{-15}~\mathrm{m~s}^{-2}/\sqrt{\mathrm{Hz}}$   
and $1~\mathrm{pm}/\sqrt{\mathrm{Hz}}$, respectively, for TianQin \cite{2016CQGra..33c5010L,Hu2018}.  
$|\tilde{C}_{1}|^{2} =1.26 \times 10^{-27}~\mathrm{Hz}^{-1}$ 
is the fractional laser frequency fluctuation (corresponding to the raw laser frequency 
noise of $10~\mathrm{Hz}/\sqrt{\mathrm{Hz}}$) \cite{2016CQGra..33c5010L,1999ApJ...527..814A}, 
which is assumed to be white in the concerned frequency range. 
Then, we find that the required time difference $\delta t_{\alpha} \leq 2.3 \times 10^{-7}$~s.

\begin{figure}[!htp]
\centering
    \subfigure[ ] {
    \begin{minipage}[t]{0.5\linewidth}
    \centering 
    \includegraphics[height=6.8cm,width=8.2cm]{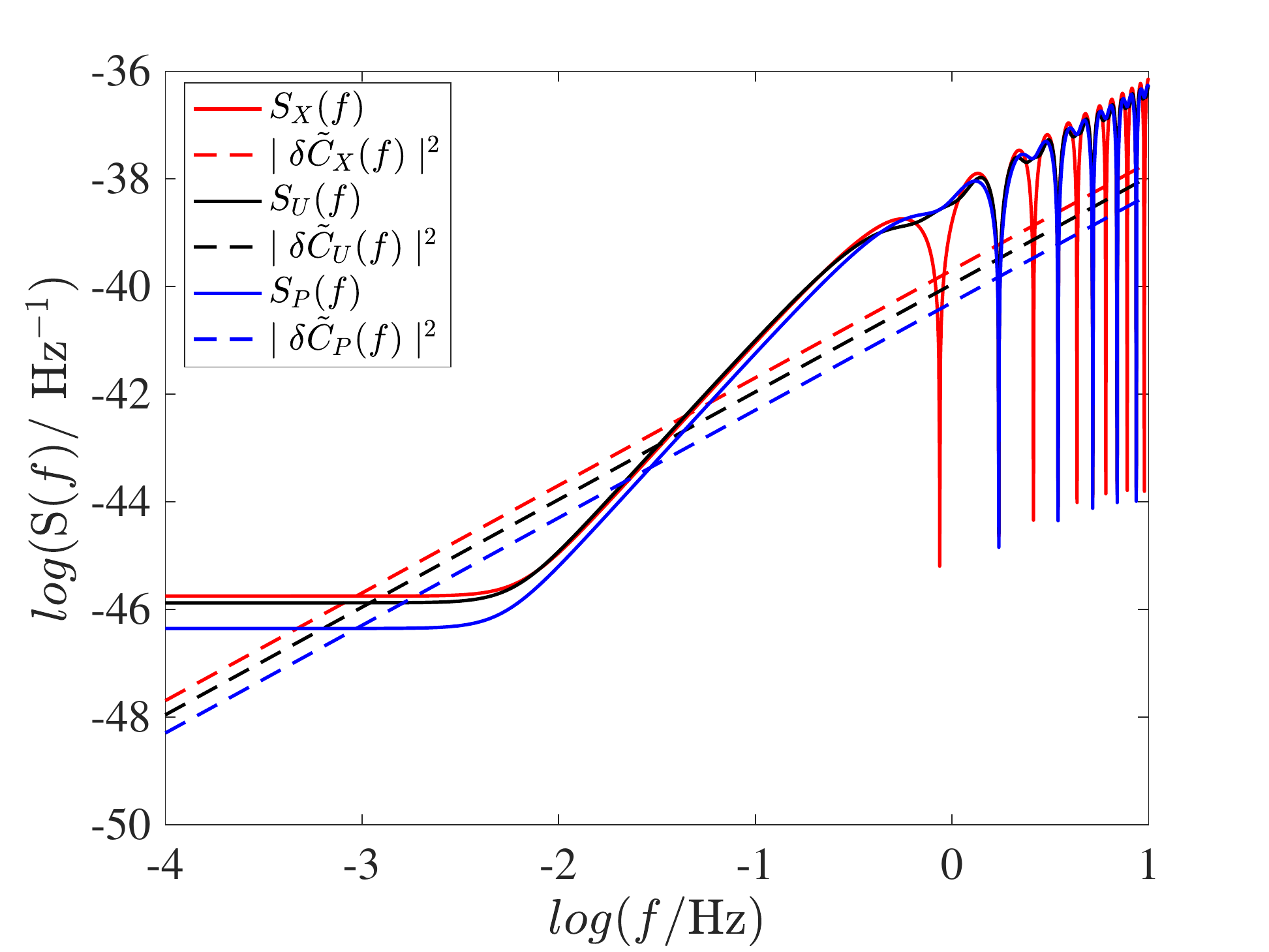} 
    \end{minipage} %
    }%
    \subfigure[ ]{
    \begin{minipage}[t]{0.5\linewidth}
    \centering 
    \includegraphics[height=6.8cm,width=8.2cm]{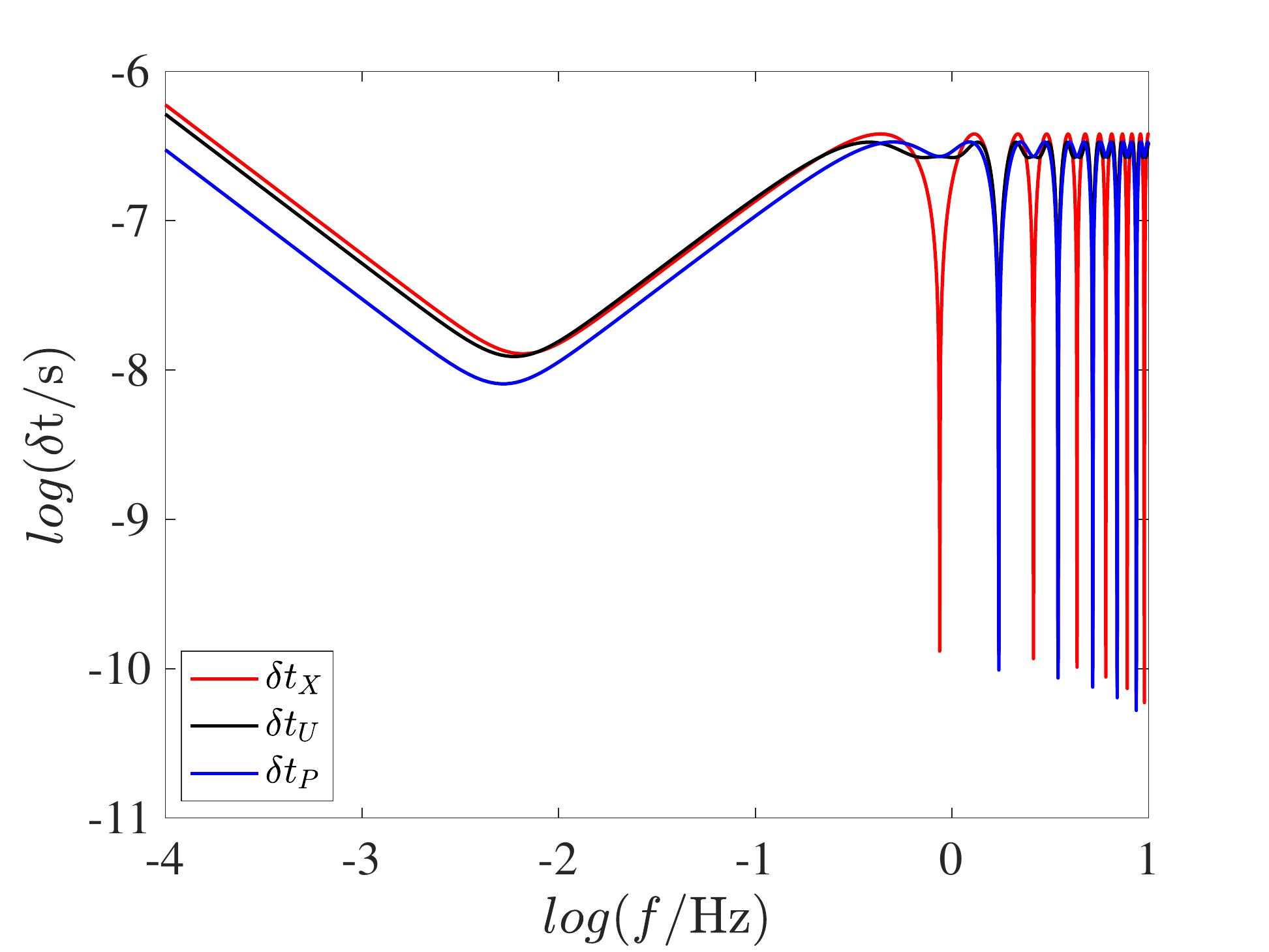} 
    \end{minipage}%
    } %
\centering  
\caption{\label{fig:secondary_noise} Left panel shows the secondary noise (solid line) and the residual laser frequency noise (dotted line) for the first-generation $X$, $U$, $P$ and $E$ combinations. The lines for $P$ and $E$ are overlapped, only the ones for $P$ are shown. Right panel shows the required time differences for the corresponding combinations. }    
\end{figure}

Similarly, for the $X$, $U$, $P$, and $E$ combinations with the corresponding secondary noise PSDs as follows, 
\begin{align}\label{eq:XUPE}
S_{X}(f)&=[4\sin^{2}(4\pi fL)+32\sin^{2}(2\pi fL)]S^{\rm accel}_{y}+16\sin^{2}(2 \pi fL)S^{\rm opt}_{y}  \,,  \\
S_{U}(f)&=[8\sin^{2}(3 \pi fL)+12\sin^{2}(2 \pi fL)+24\sin^{2}(\pi fL)]S^{\rm accel}_{y}  \nonumber  \\
&+[4\sin^{2}(3 \pi fL)+8\sin^{2}(2 \pi fL)+4\sin^{2}( \pi fL)]S^{\rm opt}_{y} \,, \\
S_{P}(f)& =[4\sin^{2}(2\pi fL)+32\sin^{2}(\pi fL)]S^{\rm accel}_{y}+[8\sin^{2}(2 \pi fL)+8\sin^{2}( \pi fL)]S^{\rm opt}_{y}  \,, \\
S_{E}(f) &= S_{P}(f)  \,.  
\end{align}
The required time differences are shown in the right panel of Fig.~\ref{fig:secondary_noise}. 

The PSDs of the secondary noises and the residual laser frequency noises 
for the $X$, $U$, $P$, and $E$ combinations are 
shown in the left panel of Fig.~\ref{fig:secondary_noise}. For the latter, the actual time differences $\delta t$  
from Fig.~\ref{fig:firstTDI} are used.  
We can see that the first-generation TDI data combinations cannot suppress the laser frequency noises 
below the secondary noises for $10^{-3} \lesssim f \lesssim 10^{-1}$~Hz; while they are effective 
at the frequencies $f \lesssim 10^{-3}$~Hz and $f \gtrsim 10^{-1}$~Hz.

\subsection{Time differences for the second-generation TDI}\label{sec:sim2}

The second-generation TDI can suppress the laser frequency noise further by 
canceling the $L_{l}$ and $L_{l}\dot{L}_{m}$ terms in Eq.~\ref{eq:fTDIale2}.  
The data combinations that have this nature are called \textit{$\dot{L}$-closed}, 
for which the interference paths can be found 
through splicing the first-generation interference paths \cite{2003PhRvD..68f1303S}. 
Taking the $\alpha$ combination as an example, 
a new path 1 (1-3-2-1-2-3-1) can be synthesized by splicing the path 1 (1-3-2-1) and 
path 2 (1-2-3-1) of the first-generation $\alpha$; a new path 2 (1-2-3-1-3-2-1) 
is in a reversed order of the satellites. 
Specifically, this process can be expressed as: 
\begin{equation} \label{inpath}
\begin{aligned}
&\overrightarrow{_{1}2_{3}1_{2}3_{1}}~\overleftarrow{_{1}2'_{3}1'_{2}3'_{1}}+\overrightarrow{_{1}3'_{2}1'_{3}2'_{1}}~\overleftarrow{_{1}3_{2}1_{3}2_{1}}\Longrightarrow
~\overrightarrow{213[3'1'2'}~\overleftarrow{312]2'1'3'}~~~(\alpha12-1) \\
=&y_{23;133'1'2'}+y_{12;33'1'2'}+y_{31;3'1'2'}+y_{32;1'2'}+y_{13;2'}+y_{21}  \\
&-y_{31}-y_{12;3}-y_{23;13}-y_{21;213}-y_{13;2'213}-y_{32;1'2'213} \,. \\
\end{aligned}
\end{equation}
Here, $y_{23;13}=y_{23}(t-L_{1}(t-L_{3}(t))-L_{3}(t))$. 
Note that we use `;' here to represent the time delay in the second generation.  
Two laser beams passing along `$\longrightarrow$' and  `$\longleftarrow$' 
interfere at time $t$. The numbers below each arrow 
represent the interference arms, the subscripts on the left hand side 
represent the indices of satellites. 
On the right hand side, $\overrightarrow{2133'1'2'}$ represents a laser beam 
passing along the arms from left to right; while, 
$\overleftarrow{3122'1'3'}$ from right to left. 
$\alpha_{1}:\overrightarrow{_{1}2_{3}1_{2}3_{1}}~\overleftarrow{_{1}2'_{3}1'_{2}3'_{1}}$  
represents the paths of two laser beams for the first-generation TDI, 
then by reversing we obtain 
$\alpha_{2}:\overrightarrow{_{1}3'_{2}1'_{3}2'_{1}}~\overleftarrow{_{1}3_{2}1_{3}2_{1}}$. 
The path $\alpha_{1}$ is spliced with the path $\alpha_{2}$ and  
result in the path $\alpha12-1$. Here, `$[~]$' indicates where we insert 
$\alpha_{2}$ into $\alpha_{1}$ \cite{2005PhRvD..72d2003V}. 

From Eq.~\ref{inpath}, we can write the time difference of the combination $\alpha12-1$ as: 
\begin{equation}\label{TDIt}
\begin{aligned}
\Delta t_{\alpha12-1}=&L_{2;133'1'2'}+L_{1;33'1'2'}+L_{3;3'1'2'}+L_{3';1'2'}+L_{1';2'}+L_{2'}\\
&-L_{3}-L_{1;3}-L_{2;13}-L_{2';213}-L_{1';2'213}-L_{3';1'2'213} \,.
\end{aligned}
\end{equation}

Similarly, the path 
$\alpha_{2}:\overrightarrow{_{1}3'_{2{\uparrow}}1'_{3}2'_{1}}~\overleftarrow{_{1}3_{2}1_{3}2_{1}}$ 
can be split at the position of  `$\uparrow$'. This results in a new path 
$\overrightarrow{_{2}1'_{3}2'_{1}}~\overleftarrow{_{1}3_{2}1_{3}2_{1}}~\overrightarrow{_{1}3'_{2}}$, 
which makes S2 as the starting and ending satellite. 
Stitching it with $\alpha_{1}$, we can obtain a new combination: 
\begin{equation}\label{TDIy}
\begin{aligned}
&\overrightarrow{_{1}2_{3}1_{2}3_{1}}~\overleftarrow{_{1}2'_{3}1'_{2}3'_{1}}+\overrightarrow{_{2}1'_{3}2'_{1}}~\overleftarrow{_{1}3_{2}1_{3}2_{1}}
~\overrightarrow{_{1}3'_{2}}\Longrightarrow~\overrightarrow{21[1'2'}~\overleftarrow{312}~\overrightarrow{3']3}\overleftarrow{2'1'3'}~~~(\alpha12-2)\\
=&y_{23;11'2'\overline{31}}+y_{12;1'2'\overline{31}}+y_{13;2'\overline{31}}+y_{21;\overline{31}}-y_{31;\overline{31}}-y_{12;\overline{1}}\\
&-y_{23}+y_{32;\overline{3'}2}+y_{31;\overline{33'}2}-y_{21;\overline{33'}2}-y_{13;2'\overline{33'}2}-y_{32;1'2'\overline{33'}2}\,.
\end{aligned}
\end{equation}
Here, following the advancement rule introduced in \cite{2005PhRvD..72d2003V}, 
we adopt the advancement indices  $\bar{l}$. For example, 
\begin{equation}
y_{32;\overline{3'}2}=y_{32}(t+L_{3'}(t-L_{2}(t))-L_{2}(t))  \,. 
\end{equation}

\begin{figure}[!htp]
\centering
    \subfigure[ ] {
    \begin{minipage}[t]{0.5\linewidth}
    \centering 
    \includegraphics[height=6.8cm,width=8.2cm]{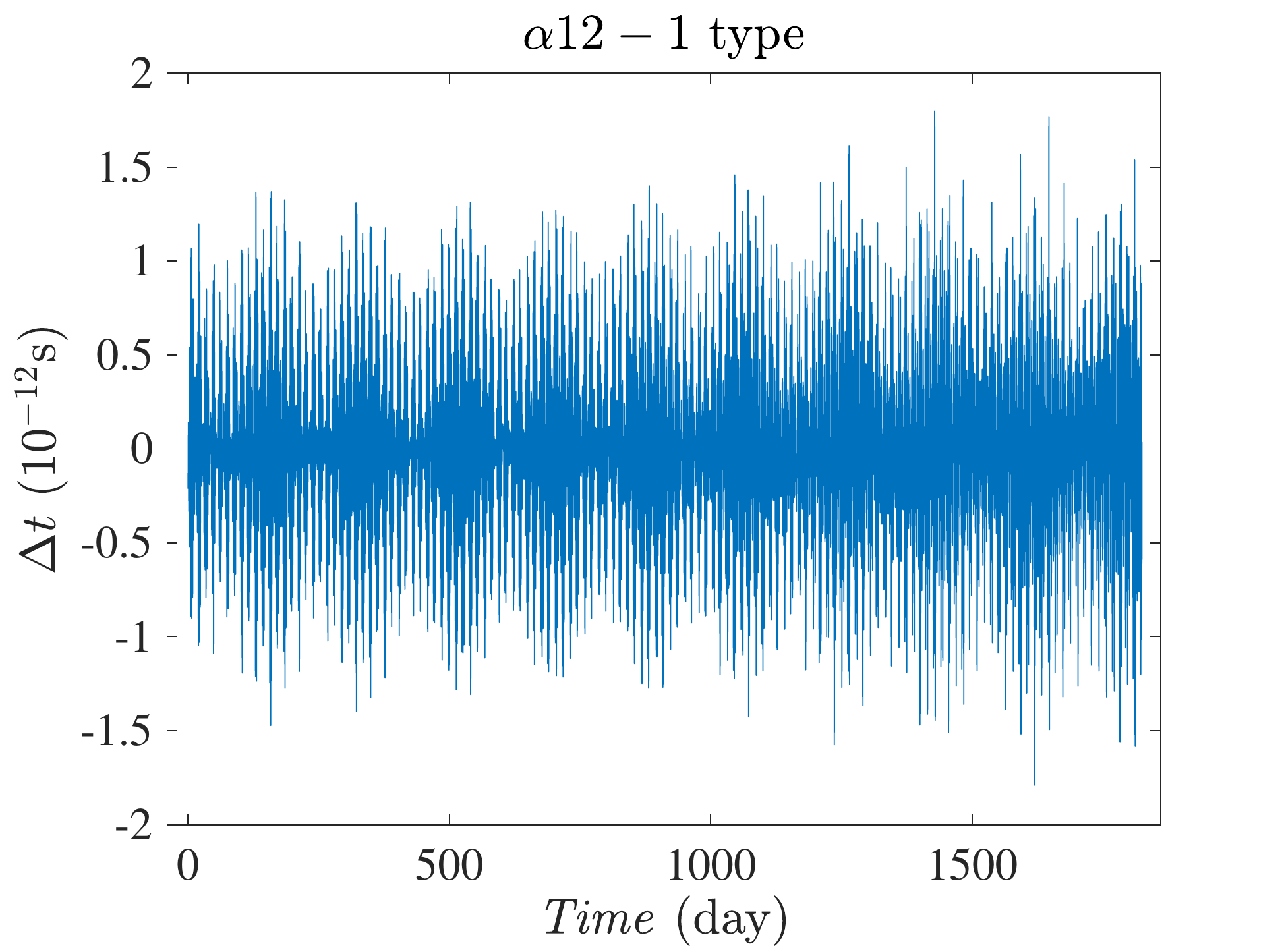} 
    \end{minipage} %
    }%
    \subfigure[ ]{
    \begin{minipage}[t]{0.5\linewidth}
    \centering 
    \includegraphics[height=6.8cm,width=8.2cm]{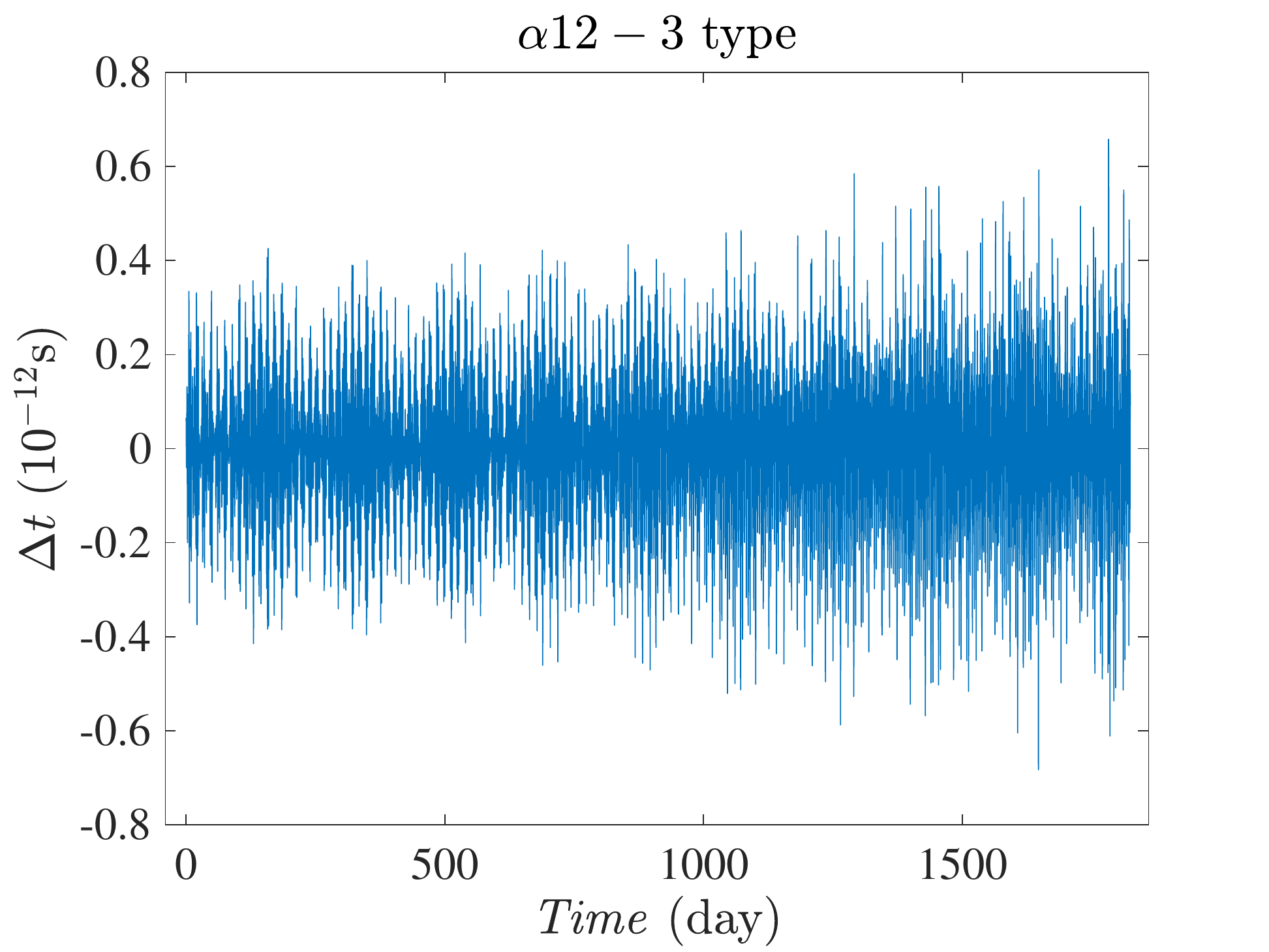} 
    \end{minipage}%
    } %
\centering  
\caption{\label{fig:alpha2} The time differences for the second-generation $\alpha$-type combinations.}    
\end{figure}

In addition, the path $\alpha_{2}$ can be split to make S3 as the starting 
and ending satellite, which gives: 
\begin{equation}
\begin{aligned}
\overrightarrow{_{1}2_{3}1_{2}3_{1}}~\overleftarrow{_{1}2'_{3}1'_{2}3'_{1}}+\overrightarrow{_{3}2'_{1}}~\overleftarrow{_{1}3_{2}1_{3}2_{1}}~
\overrightarrow{_{1}3'_{2}1'_{3}}\Longrightarrow\overrightarrow{2[2'}\overleftarrow{312}\overrightarrow{3'1']13}\overleftarrow{2'1'3'}~~~(\alpha12-3) \,.
\end{aligned}
\end{equation} 
Fig.~\ref{fig:alpha2} gives the smallest (right panel) and the largest (left panel)  time differences 
for the second-generation $\alpha$-type combinations.

For the other second-generation TDI combinations, the results are: 
\begin{equation}
\begin{aligned}
X:~&\overrightarrow{_{1}3'_{2}3_{1}2_{3}2'_{1}} ~\overleftarrow{_{1}3_{2}3'_{1}2'_{3}2_{1}}+\overrightarrow{_{1}2_{3}2'_{1}3'_{2}3_{1}} ~\overleftarrow{_{1}2'_{3}2_{1}3_{2}3'_{1}}\\
&\Longrightarrow\ \left\{\begin{aligned}\overrightarrow{3'322'[22'3'3}~\overleftarrow{2'233']33'2'2}~~~(X16-1) \,,\\
\overrightarrow{3'322'[3'3}~\overleftarrow{2'233'}~\overrightarrow{22']}~\overleftarrow{33'2'2}~~~(X16-2) \,,\\
\overrightarrow{22'3'3[22'}~\overleftarrow{33'2'2}~\overrightarrow{3'3]}~\overleftarrow{2'233'}~~~(X16-3) \,.\\
\end{aligned}
\right.
\end{aligned}
\end{equation}

The path $X_{1}:\overrightarrow{_{1}3'_{2}3_{1}2_{3}2'_{1}} ~\overleftarrow{_{1}3_{2}3'_{1}2'_{3}2_{1}}$ 
is spliced with its reversed version 
$X_{2}:\overrightarrow{_{1}2_{3}2'_{1}3'_{2}3_{1}} ~\overleftarrow{_{1}2'_{3}2_{1}3_{2}3'_{1}}$, 
resulting in the path $X16-1$. Also, the path $X_{1}:\overrightarrow{_{1}3'_{2}3_{1{\uparrow}}2_{3}2'_{1}} ~\overleftarrow{_{1}3_{2}3'_{1}2'_{3}2_{1}}$ can be split at the position of $\uparrow$, and generate the new path 
$X_{3}:\overrightarrow{22'}~\overleftarrow{33'2'2}~\overrightarrow{3'3}$, which is spliced with the path $X_{2}$ to 
form the path $X16-3$. The path $X_{2}$ can generate the new path $X_{4}:\overrightarrow{3'3}~\overleftarrow{2'233'}~\overrightarrow{22'}$, which is spliced with the path $X_{1}$ to 
form the path $X16-2$. Among these three combinations, $X16-1$ uses four directional arms 
which form a two-beam interferometers; While, $X16-2$ and $X16-3$ are four-beam interferometers. 

Similar to Eq.~\ref{inpath} and Eq.~\ref{TDIt}, we can calculate the time differences 
for the second-generation $X$ combinations for TianQin. 
The smallest (right) and largest (left) time differences are shown in Fig.~\ref{fig:X2}.  
As we will see below, the time differences of $X$ combinations are the largest comparing with in the $\alpha$, 
$U$, $P$ and $E$ combinations. 
This is because when inserting the frequency noise $C_{i}(t)$ into the combinations of `$y$' for the $X$ (the expression similar to Eq.~\ref{inpath})
and expanding in $L_{l}$ further than the canceled  $L_{l}$ and $L_{l} \dot{L}_{m}$ terms, 
the next leading terms read $L^2\ddot{L}_{m}$  
and the sums of these terms for the $X$ combinations are larger than the other combinations. 

\begin{figure}[!htp]
\centering
    \subfigure[ ] {
    \begin{minipage}[t]{0.5\linewidth}
    \centering 
    \includegraphics[height=6.8cm,width=8.2cm]{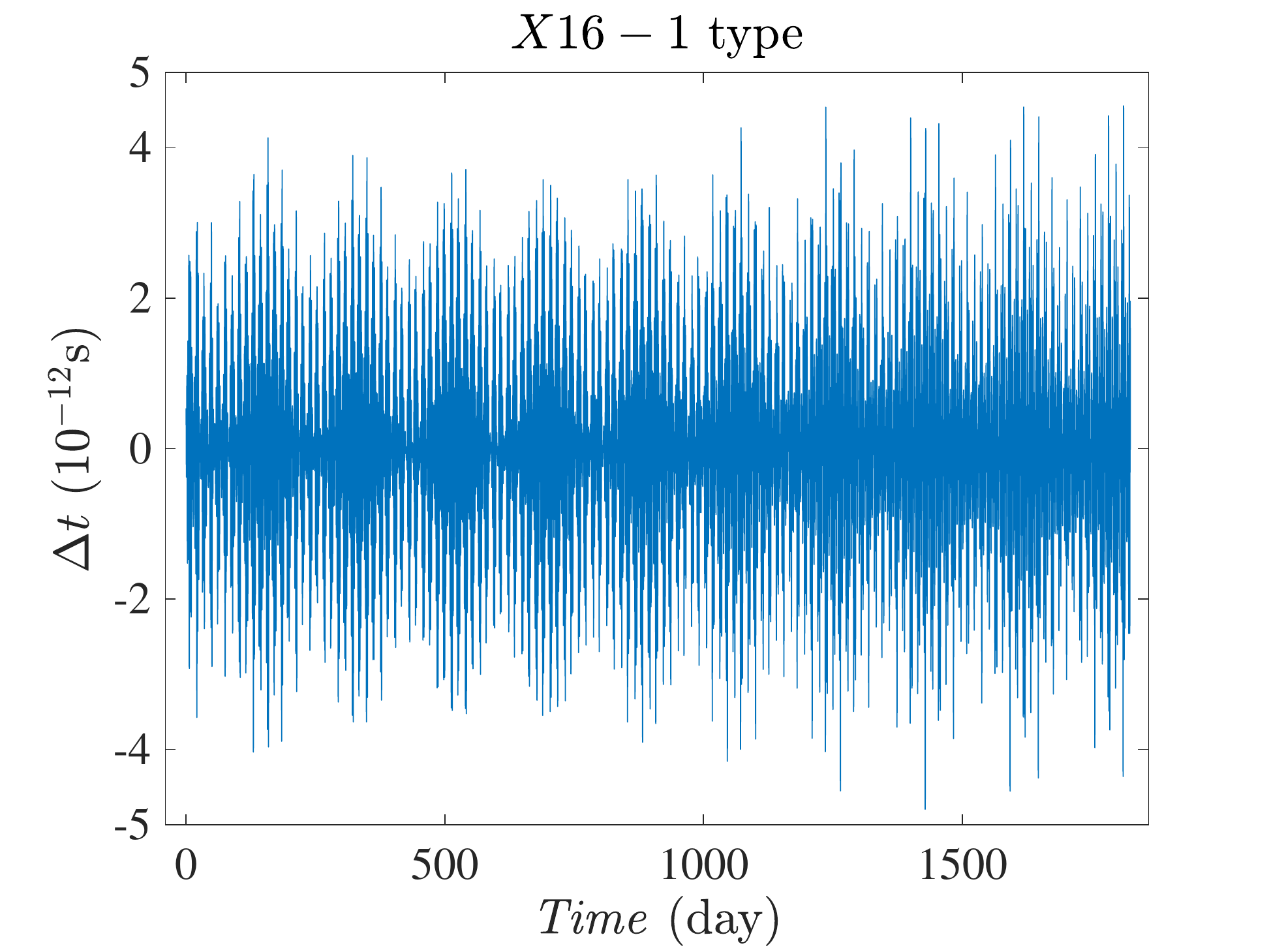} 
    \end{minipage} %
    }%
    \subfigure[ ]{
    \begin{minipage}[t]{0.5\linewidth}
    \centering 
    \includegraphics[height=6.8cm,width=8.2cm]{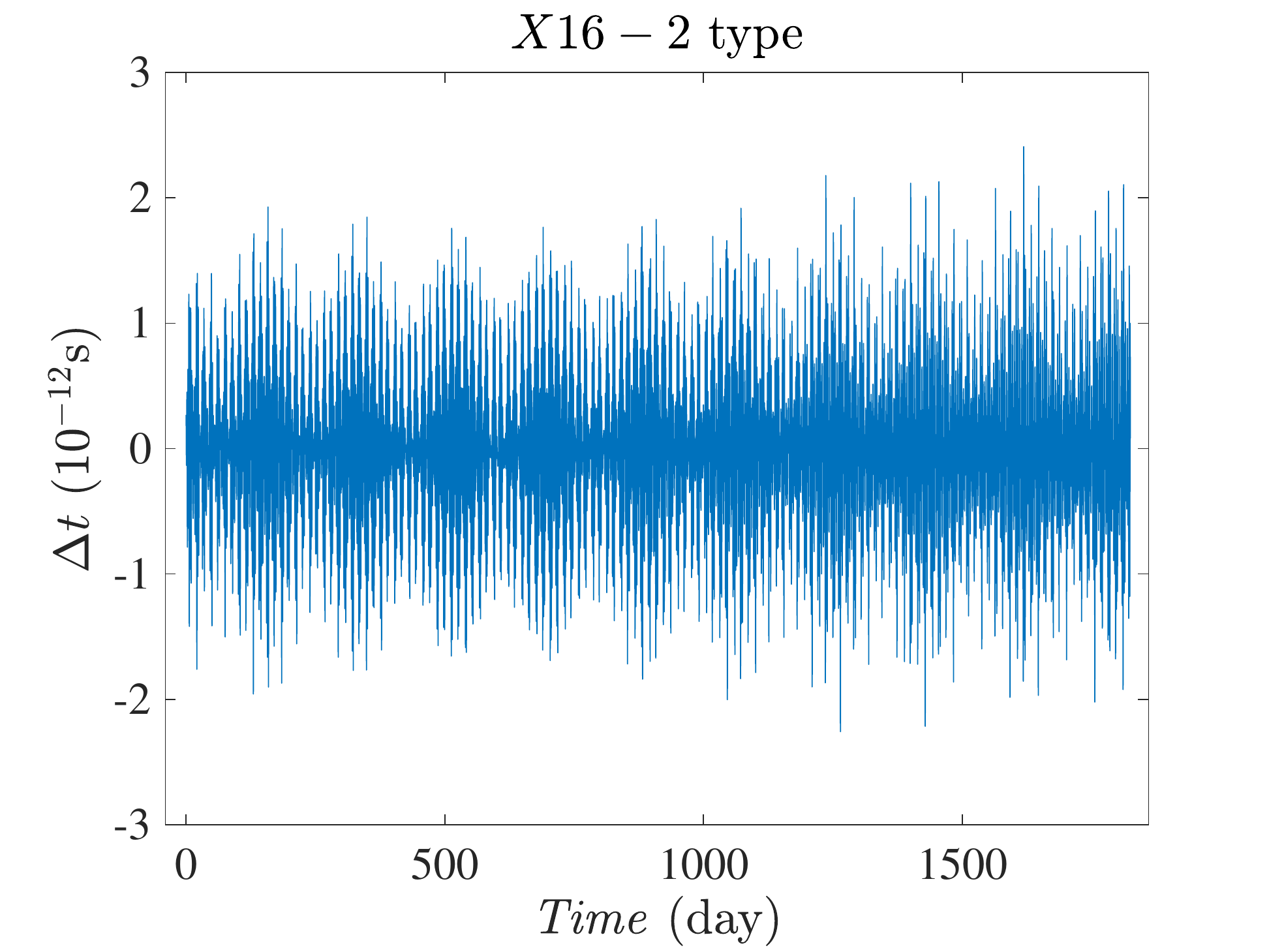} 
    \end{minipage}%
    } %
\centering  
\caption{\label{fig:X2} The time differences for the second-generation $X$-type combinations.}    
\end{figure} 

\begin{equation}
\begin{aligned}
U:~&\overrightarrow{_{3}2'_{1}3'_{2}1'_{3}1_{2}}~\overleftarrow{_{2}3'_{1}2'_{3}1'_{2}1_{3}}+\overrightarrow{_{3}1_{2}1'_{3}2'_{1}3'_{2}}~\overleftarrow{_{2}1_{3}1'_{2}3'_{1}2'_{3}}\\
&\Longrightarrow\ \left\{\begin{aligned}\overrightarrow{2'3'1'1}~\overleftarrow{3'2'1'[3'2'}~\overrightarrow{11'2'3'}~\overleftarrow{11']1}~~~(U16-1) \,,\\
\overrightarrow{2'3'1'1}~\overleftarrow{3'2'}~\overrightarrow{[11'2'3'}~\overleftarrow{11'3'2']1'1}~~~(U16-2) \,,\\
\overrightarrow{2'3'1'1[1'2'3'}~\overleftarrow{11'3'2'}~\overrightarrow{1]}~\overleftarrow{3'2'1'1}~~~(U16-3) \,.\\
\end{aligned}
\right.
\end{aligned}
\end{equation}

\begin{figure}[!htp]
\centering
    \subfigure[ ] {
    \begin{minipage}[t]{0.5\linewidth}
    \centering 
    \includegraphics[height=6.8cm,width=8.2cm]{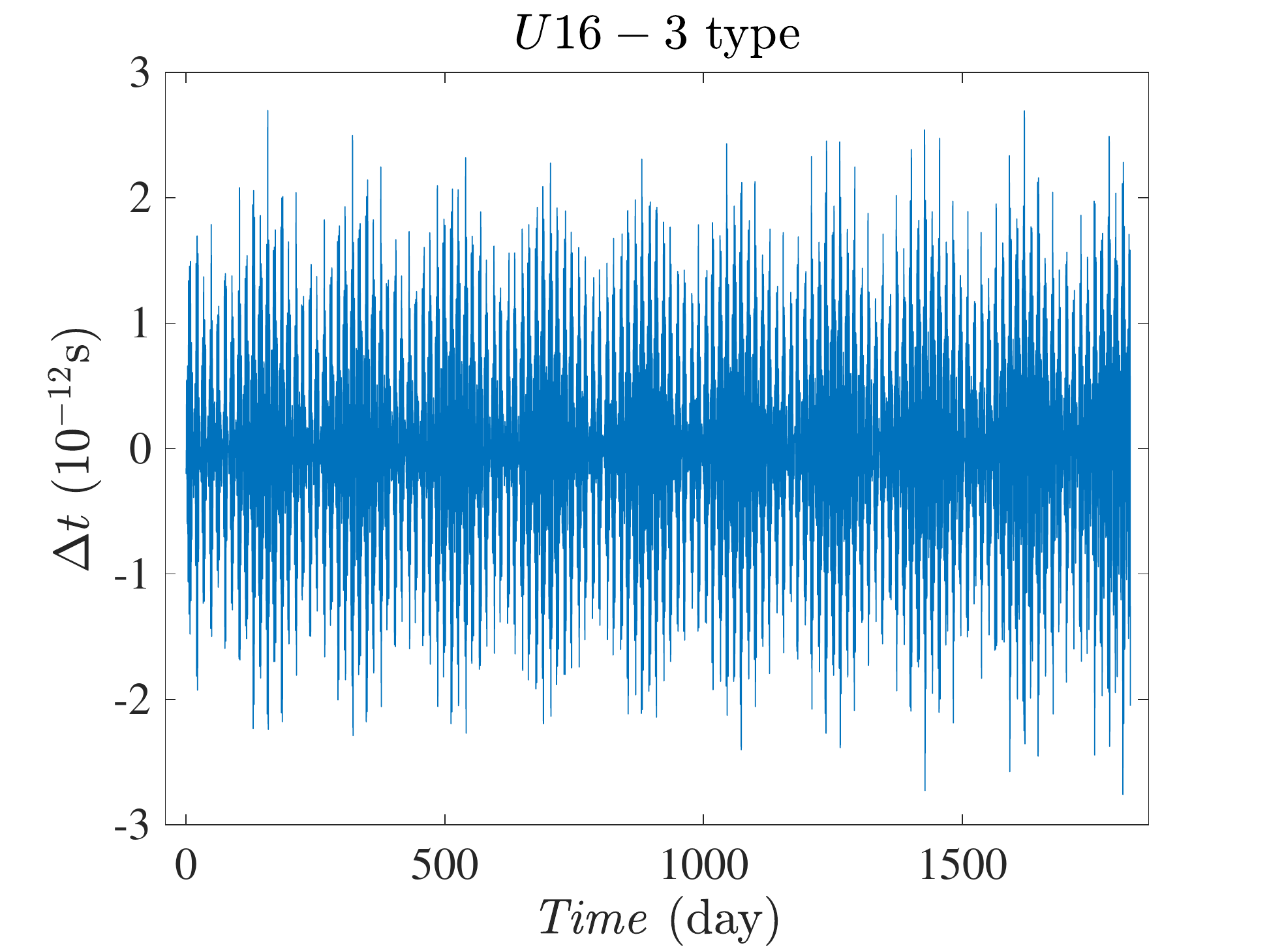} 
    \end{minipage} %
    }%
    \subfigure[ ]{
    \begin{minipage}[t]{0.5\linewidth}
    \centering 
    \includegraphics[height=6.8cm,width=8.2cm]{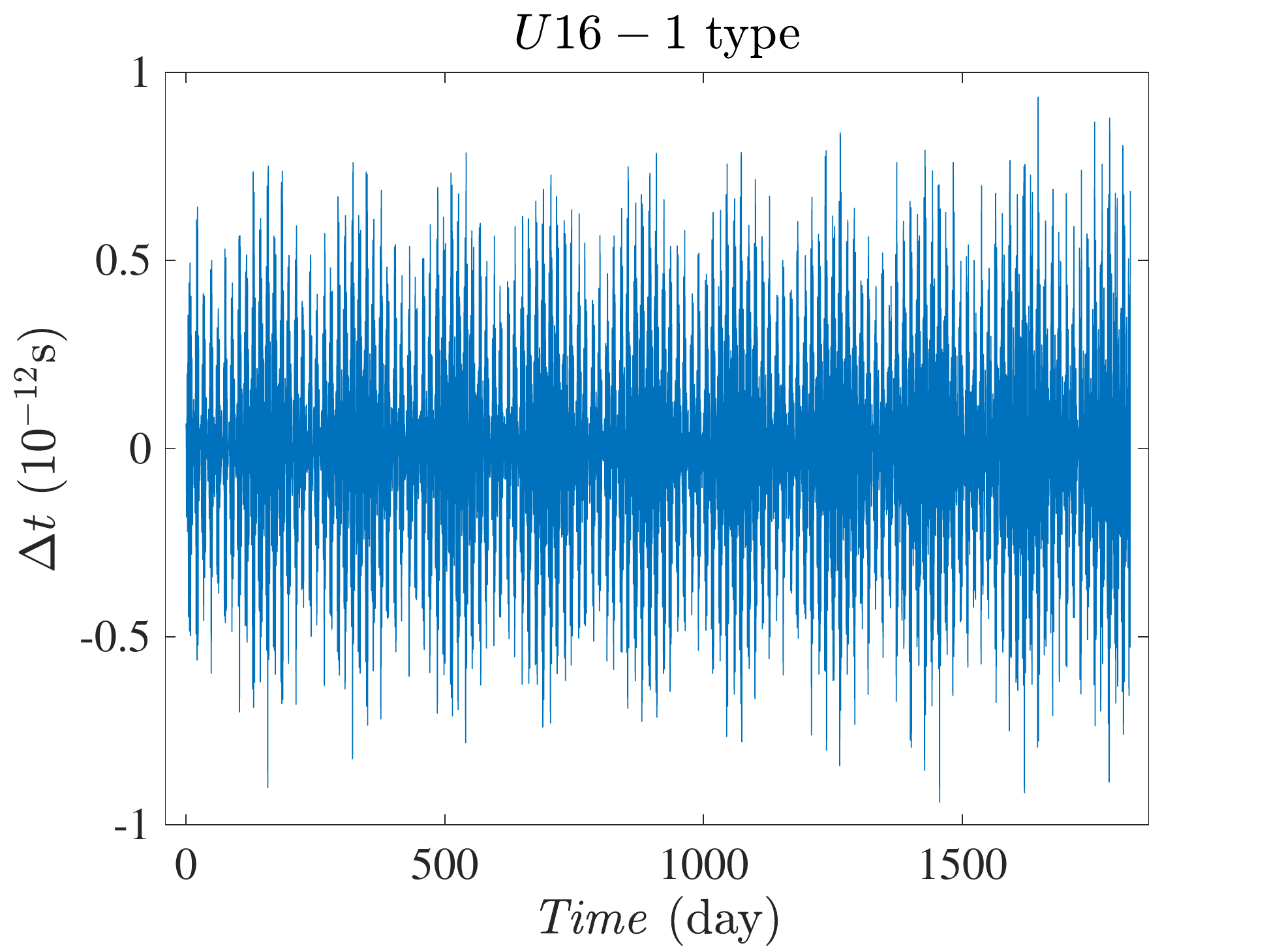} 
    \end{minipage}%
    } %
\centering  
\caption{\label{fig:U2} The time differences for the second-generation $U$-type combinations.}    
\end{figure}

The path $U_{1}:\overrightarrow{_{3}2'_{1}3'_{2}1'_{3}1_{2}}~\overleftarrow{_{2}3'_{1}2'_{3}1'_{2}1_{3}}$ is spliced with its reversed version $U_{2}:\overrightarrow{_{3}1_{2}1'_{3}2'_{1}3'_{2}}~\overleftarrow{_{2}1_{3}1'_{2}3'_{1}2'_{3}}$, resulting in the path $U16-2$. Also, the path $U_{2}:\overrightarrow{_{3}1_{2{\uparrow}}1'_{3}2'_{1}3'_{2}}~\overleftarrow{_{2}1_{3}1'_{2{\uparrow}}3'_{1}2'_{3}}$ can be split at  the position of $\uparrow$, and generate the new path $U_{3}:~\overrightarrow{1'2'3'}~\overleftarrow{11'3'2'}~\overrightarrow{1}$ and $U_{4}:~\overleftarrow{3'2'}~\overrightarrow{11'2'3'}~\overleftarrow{11'}$. $U_{3}$ is spliced with the path $U_{1}$ to form the path  $U16-3$. $U16-1$ is obtained by $U_{1}$ and $U_{4}$. All the $U$-type combinations are four-beam interferometers. 
Here, four directional arms ${2',3',1',1}$ are used. 
Fig.~\ref{fig:U2} gives the smallest (right panel) and the largest (left panel)  time differences 
for the second-generation $U$-type combinations.

\begin{equation}
\begin{aligned}
P:~&\overrightarrow{_{1}3'_{2}1'_{3}1_{2}}~\overleftarrow{_{2}3'_{1}}~\overrightarrow{_{1}2_{3}}~\overleftarrow{_{3}1'_{2}1_{3}2_{1}}+
\overrightarrow{_{1}2_{3}1_{2}1'_{3}}~\overleftarrow{_{3}2_{1}}~\overrightarrow{_{1}3'_{2}}~\overleftarrow{_{2}1_{3}1'_{2}3'_{1}}\\
&\Longrightarrow
\left\{\begin{aligned}\overrightarrow{3'1'1[1'}~\overleftarrow{2}~\overrightarrow{3'}~\overleftarrow{11'3'}~\overrightarrow{21]}~\overleftarrow{3'}~\overrightarrow{2}~\overleftarrow{1'12}~~~(P16-1) \,,\\
\overrightarrow{211'[1}~\overleftarrow{3'}~\overrightarrow{2}~\overleftarrow{1'12}~\overrightarrow{3'1']}~\overleftarrow{2}~\overrightarrow{3'}~\overleftarrow{11'3'}~~~(P16-2) \,,\\
\overrightarrow{211'}~\overleftarrow{2}~\overrightarrow{[3'1'1}~\overleftarrow{3'}~\overrightarrow{2}~\overleftarrow{1'12]}~\overrightarrow{3'}~\overleftarrow{11'3'}~~~(P16-3) \,.\\
\end{aligned}
\right.
\end{aligned}
\end{equation}

\begin{figure}[!htp]
\centering
    \subfigure[ ] {
    \begin{minipage}[t]{0.5\linewidth}
    \centering 
    \includegraphics[height=6.8cm,width=8.2cm]{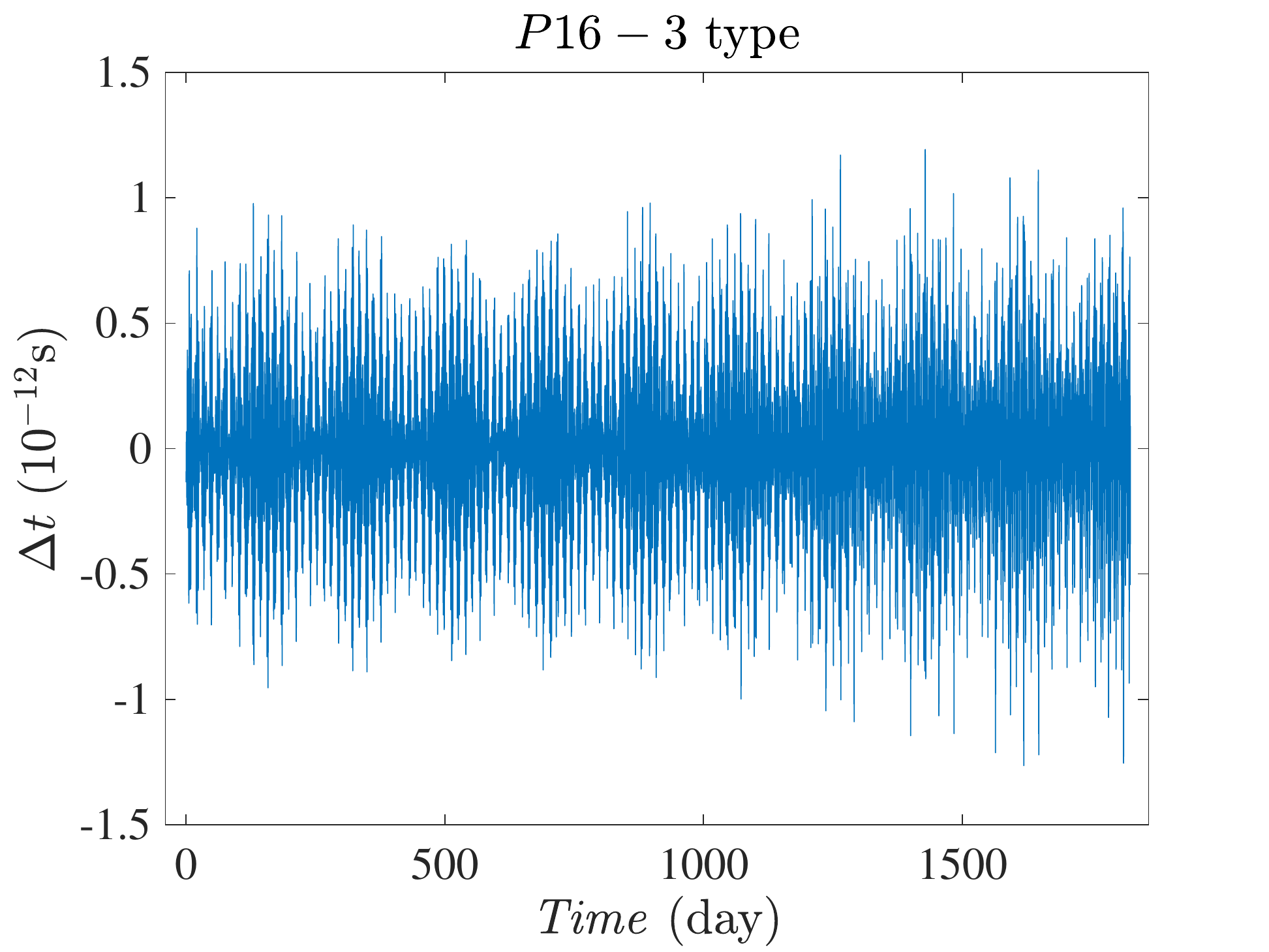} 
    \end{minipage} %
    }%
    \subfigure[ ]{
    \begin{minipage}[t]{0.5\linewidth}
    \centering 
    \includegraphics[height=6.8cm,width=8.2cm]{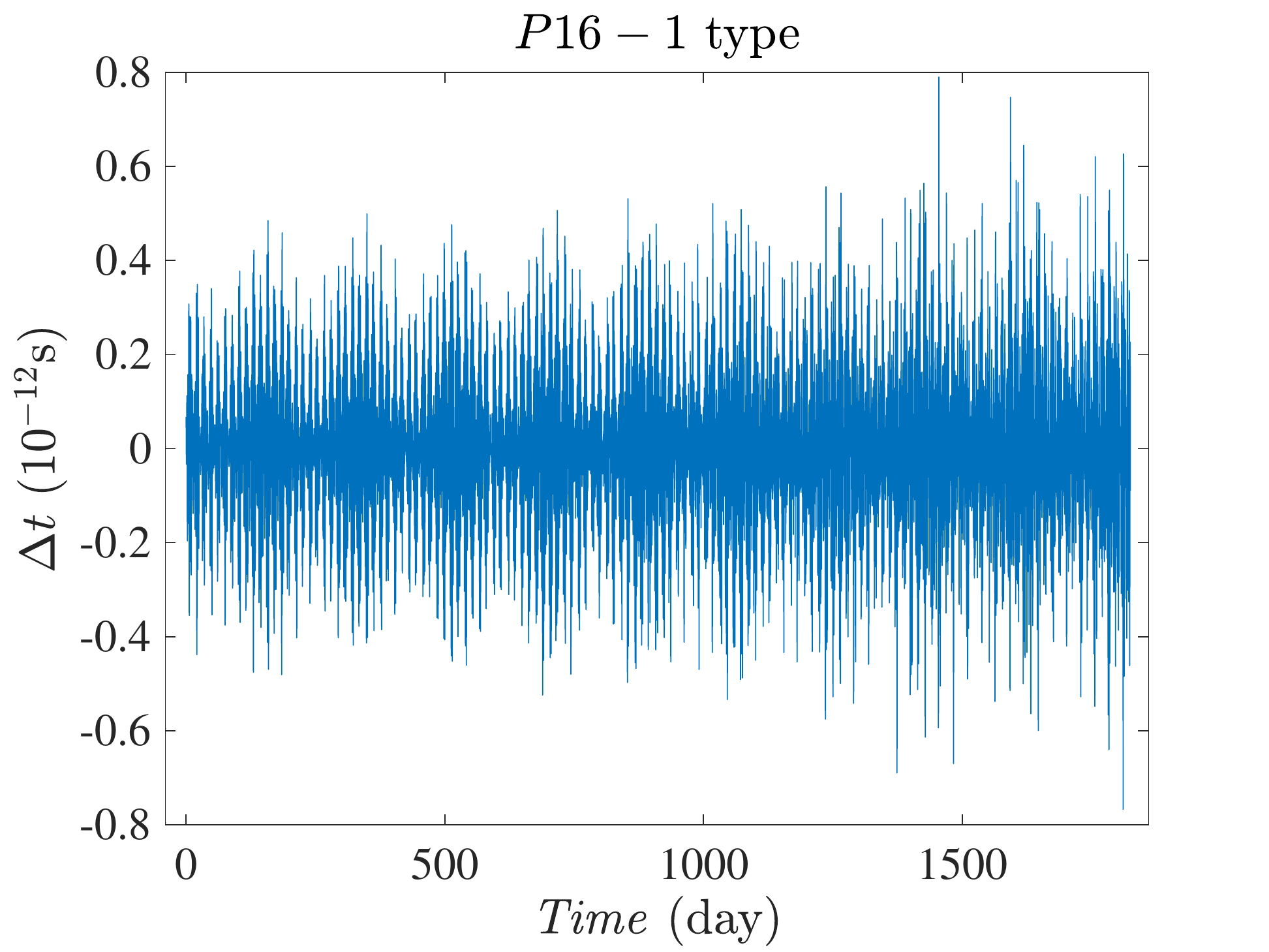} 
    \end{minipage}%
    } %
\centering      
\caption{\label{fig:P2} The time differences for the second-generation $P$-type combinations.}    
\end{figure}

The path $P_{1}:~\overrightarrow{_{1}3'_{2}1'_{3}1_{2}}~\overleftarrow{_{2}3'_{1}}~\overrightarrow{_{1}2_{3}}~\overleftarrow{_{3}1'_{2}1_{3}2_{1}}$ is spliced with its reversed version 
$P_{2}:~\overrightarrow{_{1}2_{3}1_{2}1'_{3}}~\overleftarrow{_{3}2_{1}}~\overrightarrow{_{1}3'_{2}}~\overleftarrow{_{2}1_{3}1'_{2}3'_{1}}$, resulting in the path $P16-3$. Also, the path $P_{2}:~\overrightarrow{_{1}2_{3}1_{2{\uparrow}}1'_{3}}~\overleftarrow{_{3}2_{1}}~\overrightarrow{_{1}3'_{2}}~\overleftarrow{_{2}1_{3}1'_{2}3'_{1}}$ can be split at the position of $\uparrow$, and generate the new path $P_{3}:~\overrightarrow{1'}~\overleftarrow{2}~\overrightarrow{3'}~\overleftarrow{11'3'}~\overrightarrow{21}$, which is spliced with the path $P_{1}$ to form $P16-1$.   From $P_{1}$ we can derive  $P_{4}:~ \overrightarrow{1}~\overleftarrow{3'}~\overrightarrow{2}~\overleftarrow{1'12}~\overrightarrow{3'1'}$, which is spliced with the path $P_{2}$ to form $P16-2$.
Fig.~\ref{fig:P2} gives the smallest (right panel) and the largest (left panel)  time differences 
for the second-generation $P$-type combinations.

\begin{equation}
\begin{aligned}
E:~&\overrightarrow{_{2}1'_{3}1_{2}3_{1}}~\overleftarrow{_{1}2'_{3}1'_{2}1_{3}}~\overrightarrow{_{3}2'_{1}}~\overleftarrow{_{1}3_{2}}+
\overrightarrow{_{2}3_{1}}~\overleftarrow{_{1}2'_{3}}~\overrightarrow{_{3}1_{2}1'_{3}2'_{1}}~\overleftarrow{_{1}3_{2}1_{3}1'_{2}}\\
&\Longrightarrow
\left\{\begin{aligned}\overrightarrow{1[1'13}~\overleftarrow{2'1'1}~\overrightarrow{2'}~\overleftarrow{3]}~\overrightarrow{1'2'}~\overleftarrow{311'}~\overrightarrow{3}~\overleftarrow{2'}~~~(E16-1) \,,\\
\overrightarrow{1'[11'2'}~\overleftarrow{311'}~\overrightarrow{3}~\overleftarrow{2']}~\overrightarrow{13}~\overleftarrow{2'1'1}~\overrightarrow{2'}~\overleftarrow{3}~~~(E16-2) \,,\\
\overrightarrow{11'2'}~\overleftarrow{[3}~\overrightarrow{1'13}~\overleftarrow{2'1'1}~\overrightarrow{2']}~\overrightarrow{311'}~\overrightarrow{3}~\overleftarrow{2'}~~~(E16-3) \,.\\
\end{aligned}
\right.
\end{aligned}
\end{equation}

\begin{figure}[!htp]
\centering
    \subfigure[ ] {
    \begin{minipage}[t]{0.5\linewidth}
    \centering 
    \includegraphics[height=6.8cm,width=8.2cm]{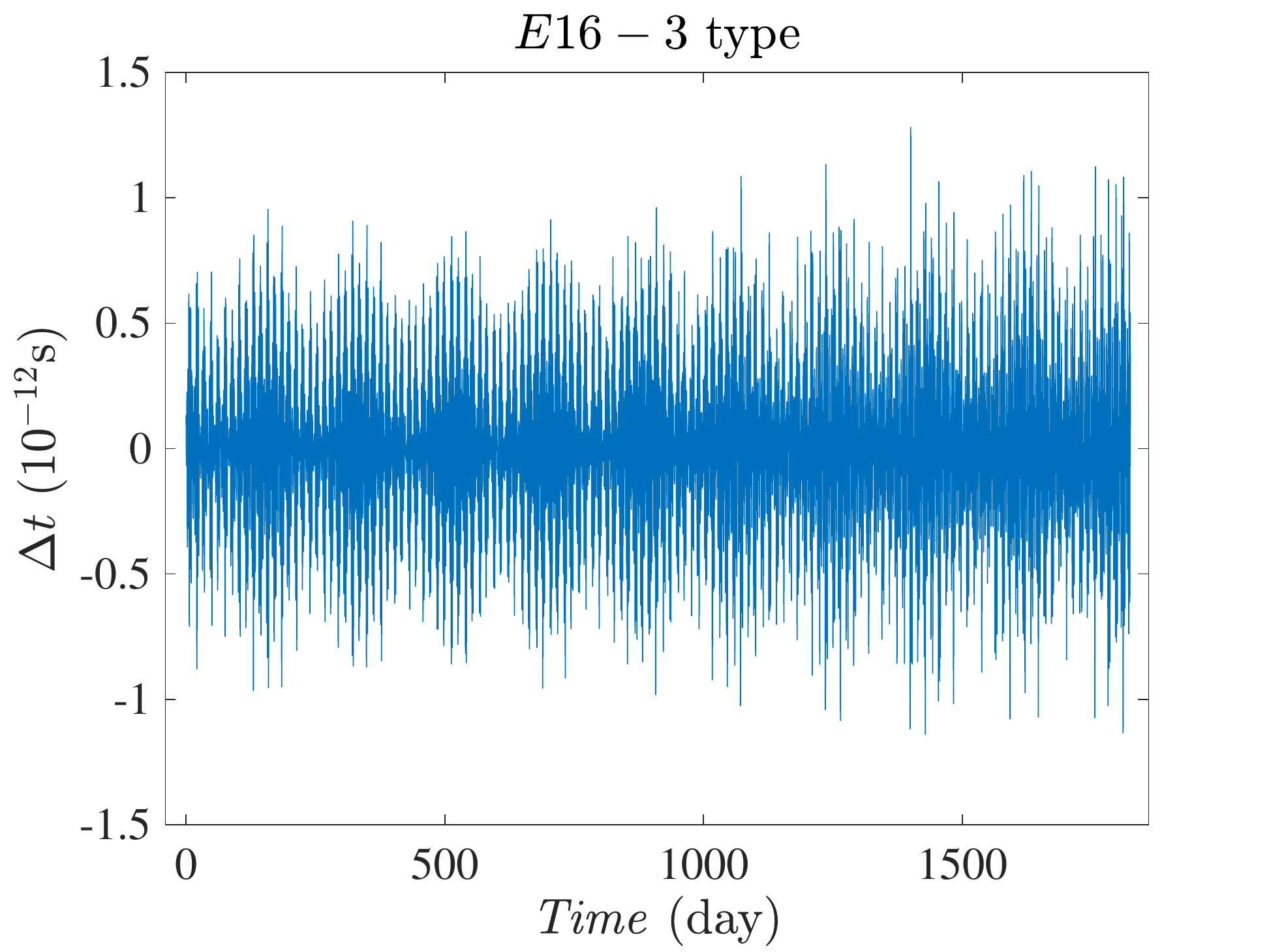} 
    \end{minipage} %
    }%
    \subfigure[ ]{
    \begin{minipage}[t]{0.5\linewidth}
    \centering 
    \includegraphics[height=6.8cm,width=8.2cm]{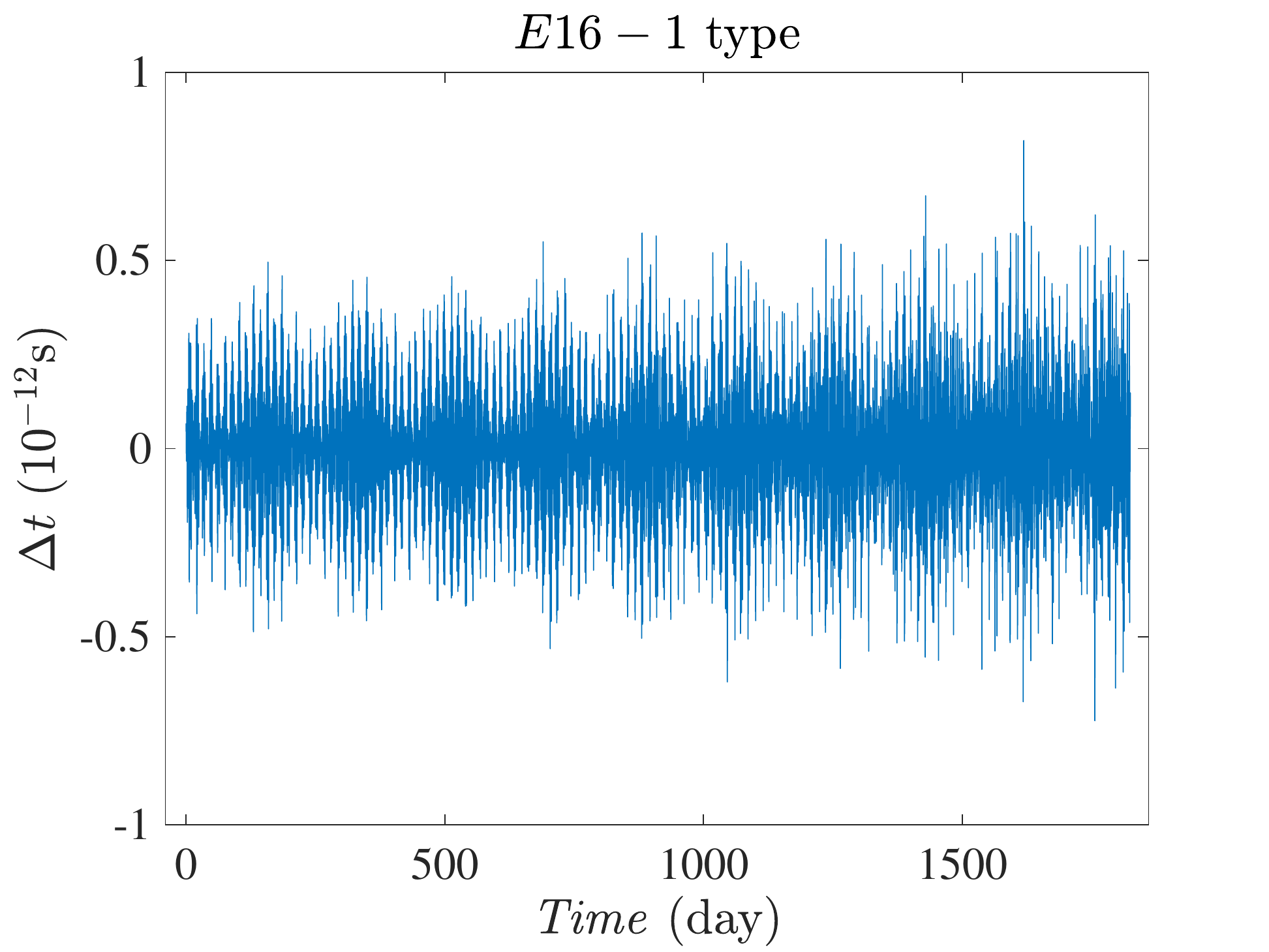} 
    \end{minipage}%
    } %
\centering  
\caption{\label{fig:E2} The time differences for the second-generation $E$-type combinations.}    
\end{figure}

By reversing the interference arm of $E_{1}:\overrightarrow{_{2}1'_{3}1_{2}3_{1}}~\overleftarrow{_{1}2'_{3}1'_{2}1_{3}}~\overrightarrow{_{3}2'_{1}}~\overleftarrow{_{1}3_{2}}$, we can obtain $E_{2}:\overrightarrow{_{2}3_{1}}~\overleftarrow{_{1}2'_{3}}~\overrightarrow{_{3}1_{2}1'_{3}2'_{1}}~\overleftarrow{_{1}3_{2}1_{3}1'_{2}}$. 
$E_{2}:\overrightarrow{_{2}3_{1}}~\overleftarrow{_{1}2'_{3{\uparrow}}}~\overrightarrow{_{3}1_{2}1'_{3}2'_{1}}~\overleftarrow{_{1}3_{2}1_{3}1'_{2}}$ can be split at the position of $\uparrow$, and generate the new path
$E_{3}:~\overrightarrow{11'2'}~\overleftarrow{311'}~\overrightarrow{3}~\overleftarrow{2'}$, which is spliced with the path $E_{1}$ to form the path of $E16-1$ and $E16-2$. Also, the path $E_{1}$ can generate the new path of $E_{4}:\overleftarrow{3}~\overrightarrow{1'13}~\overleftarrow{2'1'1}~\overleftarrow{2'}$, which is spliced with the path of $E_{3}$ to form the path $E16-3$. The combinations of $E$-type and $P$-type both form eight-beam interferometers. 
Fig.~\ref{fig:E2} gives the smallest (right panel) and the largest (left panel)  time differences 
for the second-generation $E$-type combinations.

\begin{figure}[!htp]
\centering
    \subfigure[ ] {
    \begin{minipage}[t]{0.5\linewidth}
    \centering 
    \includegraphics[height=6.8cm,width=8.2cm]{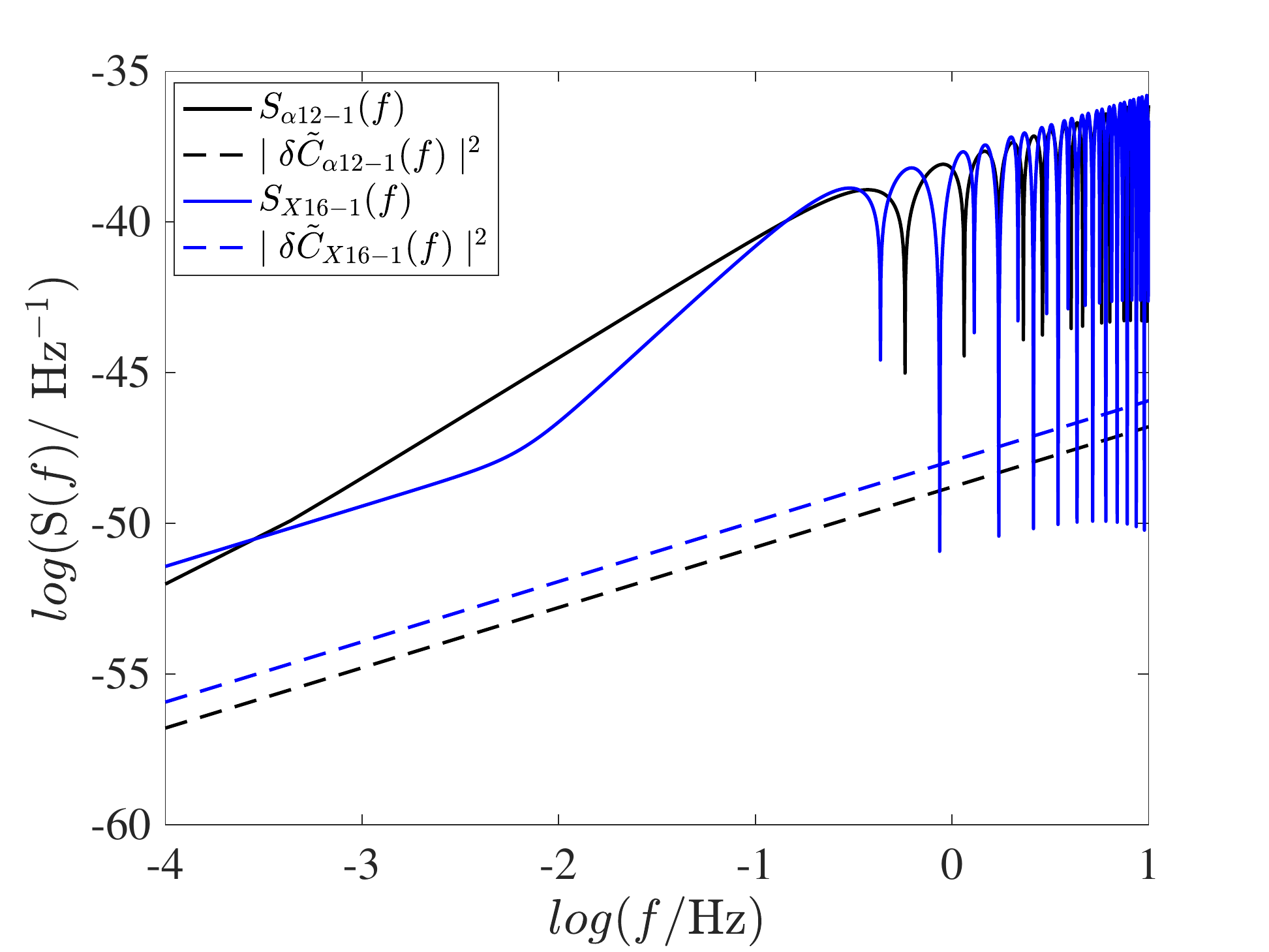} 
    \end{minipage} %
    }%
    \subfigure[ ]{
    \begin{minipage}[t]{0.5\linewidth}
    \centering 
    \includegraphics[height=6.8cm,width=8.2cm]{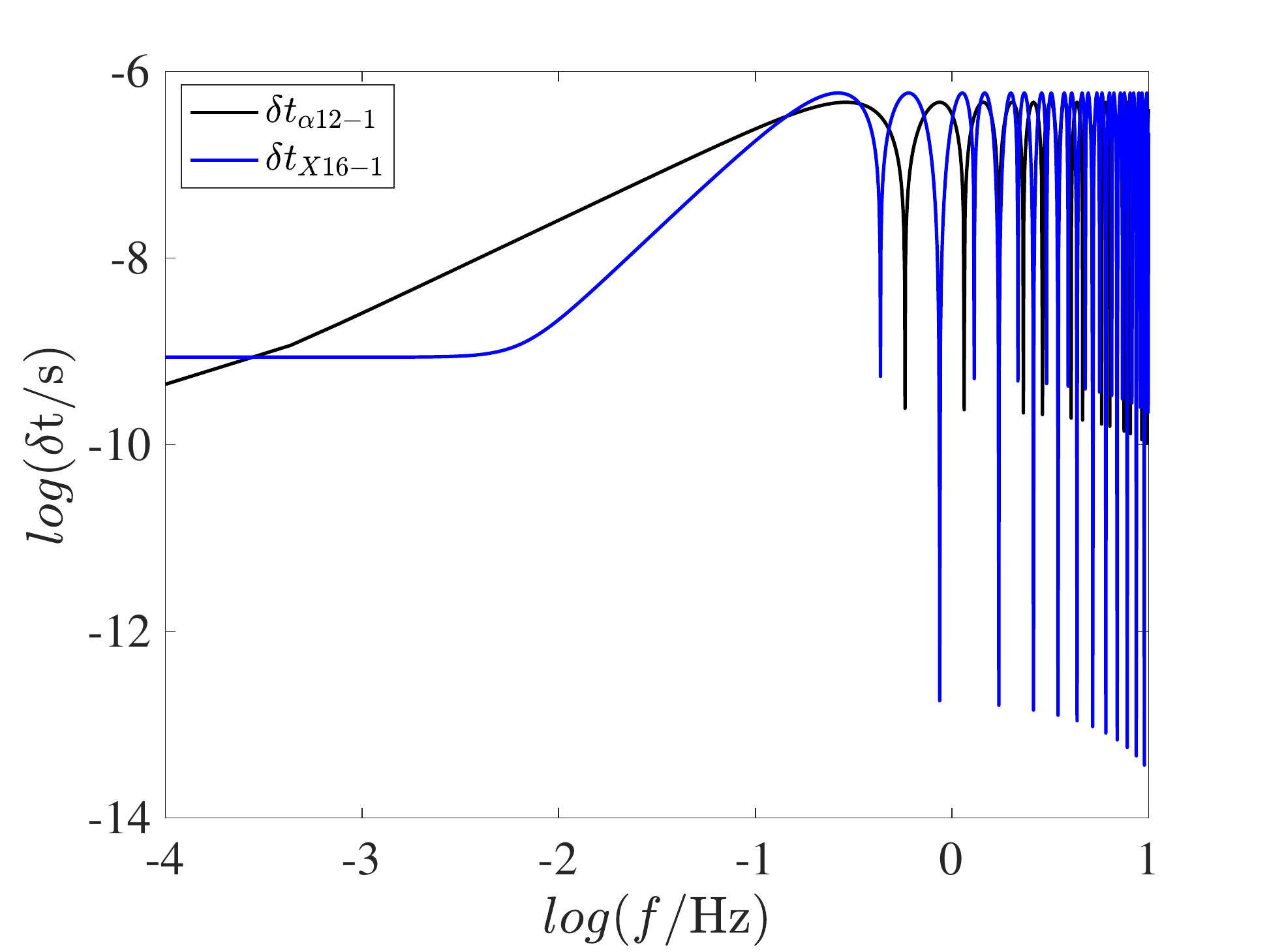} 
    \end{minipage}%
    } %
\centering  
\caption{\label{fig:secondary_noise_second} Left panel shows the secondary noise (solid line) and the residual laser frequency noise (dotted line) for the second-generation $\alpha 12-1$ and $X 16-1$ combinations. Right panel shows the required time differences for the corresponding combinations. 
}    
\end{figure}

As we can see from the results above, the eight-beam interferometers are 
better than the other second-generation TDI combinations. 
The time differences of all the second-generation TDI data combinations are below $5\times 10^{-12}$~s. 
Similar to Fig.~\ref{fig:secondary_noise}, the results for the second-generation combinations, 
taking $\alpha12-1$ and $X16-1$ as examples, are shown in Fig.~\ref{fig:secondary_noise_second}. Here,  
the secondary noise PSDs $S_{\alpha12-1}(f) = 4\sin^{2}(3\pi f L)S_{\alpha}(f)$ 
and $S_{X16-1}(f) = 4\sin^{2}(4\pi f L)S_{X}(f)$ \cite{2004PhRvD..70b2003K} are adopted. 
We can see that the time differences of the second-generation data combinations ($\sim 10^{-12}$~s) 
are substantially lower than the requirements ($>10^{-10}$~s), thus they are 
guaranteed to suppress the laser frequency noise 
well below the secondary noises for TianQin in the concerned frequencies.

\section{Conclusion and discussions}\label{sec:sum} 

Using the numerically optimized orbit that shows realistic features of the satellite constellation, 
we investigated the time differences of the symmetric interference paths, 
as a measure of residual laser frequency noise, of various first-  
and second-generation TDI data combinations for TianQin. 
We found that while the second-generation TDI with a typical time differences of $10^{-12}$~s  
is guaranteed to be valid for laser frequency noise suppression, the first-generation TDI is possible 
to be competent for GW signals at frequencies $\lesssim 10^{-3}$~Hz and  $\gtrsim 10^{-1}$~Hz 
(given the raw laser frequency noise of $10~\mathrm{Hz}/\sqrt{\mathrm{Hz}}$), 
which cover the coalescence of the SMBHB with a redshifted total mass $>10^{6}~M_{\odot}$ \cite{2019PhRvD..99l3002F} 
for the former or the inspiral of the stellar-mass black hole binary 
(e.g., GW150914) $\lesssim 1$~yr before its final merger \cite{PhysRevD.101.103027} for the latter. 
The first-generation TDI combinations will become fully useful with improved stabilization of the raw 
laser frequency noise down to $\approx 1~\mathrm{Hz}/\sqrt{\mathrm{Hz}}$ in $10^{-3} -10^{-2}$~Hz. 
This further stabilization of laser frequency can be possibly achieved through suppressing the fluctuation of 
the Pound-Drever-Hall error signal offset, 
thermal stabilization at the zero-crossing temperature of the optical bench coefficient of
thermal expansion, upgrading the automatic functions of the digital controller,
etc. \cite{Luo2016,Nicklaus2017,Ming_2020}. 
The first-generation TDI combinations, when implemented, will simplify the data analysis procedure and reduce the
computational cost, since they employ half of the single-link data than the corresponding second-generation ones.

The current work serves as our first step towards building an end-to-end TDI simulation for TianQin. 
Here, we only discussed the inter-satellite measurements `$y$'. 
In reality, when each laser on the two optical benches housed in a satellite are unlocked 
or the acceleration noises of the optical benches are concerned, 
the intra-satellite measurements `$z$' need to be included in the simulation, 
although which are not subjected to the orbital effects of the satellites discussed here. 
The simulated TDI data combinations, realistic noises from various subsystems, interplanetary and 
relativistic effects on the optical path, etc. will be taken as the input of subsequent GW data analysis for various 
astrophysical sources. 
These will be investigated in our future study.

\begin{acknowledgments}

Y.W. is supported by the National Natural Science Foundation of China (NSFC) 
under Grants No. 11973024 and No. 11690021, and 
Guangdong Major Project of Basic and Applied Basic Research (Grant No. 2019B030302001). 
The contribution of S.C.H. to this paper is supported by NSFC under Grants No. 11873098. 
X.Z. is supported by NSFC Grant No. 11805287. 
W.S. acknowledges the support from the NSFC under Grant No. 11803008 
and National Key Research and Development Program of China (Grant No. 2020YFC2201201).  
We thank the anonymous referee for helpful comments and suggestions. 

\end{acknowledgments}


\end{document}